\documentclass[12pt]{iopart}
\usepackage{iopams}  

\usepackage{graphicx}  
\usepackage{dcolumn}   
\usepackage{bm}        
\usepackage{xspace}
\usepackage{xcolor}
\usepackage{float}
\usepackage{subcaption}
\usepackage{enumitem}

\usepackage [T1]{fontenc}
\usepackage [utf8]{inputenc}
\usepackage[paperwidth=210mm,paperheight=297mm,centering,hmargin=2cm,vmargin=2.5cm]{geometry}
\usepackage{multirow}

\usepackage[square,numbers,sort&compress]{natbib}
\usepackage[colorlinks]{hyperref}
\newcommand{\text}{\mathrm}
\newcommand{\eqref}[1]{(\ref{#1})}
\newcommand{\be}{\begin{equation}}
	\newcommand{\ee}{\end{equation}}
\newcommand{\bea}{\begin{eqnarray}}
	\newcommand{\eea}{\end{eqnarray}}

\graphicspath{{./Figures/}{./}}

\begin{document}

\title{Critical quench dynamics of Wegner's $\mathbb{Z}_2$ gauge model:\\ a geometric perspective}

\author{Ramgopal Agrawal$^{1,2}$, Leticia F. Cugliandolo$^{2}$, and Marco Picco$^2$}
\address{$^1$Dipartimento di Fisica, Sapienza Universit\'a di Roma, P.le A. Moro 2, I-00185, Roma, Italy}
\address{$^2$Sorbonne Universit\'e, Laboratoire de Physique Th\'eorique et Hautes Energies, \\CNRS UMR 7589, 4 Place Jussieu, 75252 Paris Cedex 05, France}

\begin{abstract}
\noindent
Wegner's $\mathbb{Z}_2$ gauge model is the earliest formulation of pure lattice gauge theory and predicts the topological nature of the \textit{confinement--deconfinement} transition. In three dimensions ($D=3$), the equilibrium critical behavior of the model is understood in terms of \textit{geometrically} defined objects, namely loop excitations and Fortuin--Kasteleyn (FK) clusters. This work investigates the critical quench dynamics of this model from a geometric perspective, following quenches from both a high-temperature percolation phase and the zero-temperature ground state. Using time-dependent finite-size scaling analysis, we find that the critical non-equilibrium relaxation of the percolation order parameter is governed by a dynamical exponent $z_{\rm p} \simeq 2.6$, consistent with that associated with the energy density, $z_{\rm c}$. Importantly, the value of $z_{\rm p}$ is robust with respect to the initial quench condition and the choice of geometrical objects. Furthermore, we provide a detailed characterization of the kinetics of different geometrical objects during the evolution from the percolation phase. Notably, we observe that the quench dynamics obeys dynamic scaling in terms of a growing lengthscale, $\xi_{\rm p}(t) \sim t^{1/z_{\rm p}}$, despite the absence of a local order parameter.
\end{abstract}

\vspace{1pc}



\maketitle

\pagestyle{plain}
\tableofcontents

\section{Introduction}
\label{S1}

One of the ubiquitous features of lattice gauge theories  is their \textit{local gauge invariance}, which precludes the existence of a conventional local order parameter. A prototypical example is Wegner's $\mathbb{Z}_2$ three-dimensional gauge model~\cite{10.1063/1.1665530}, which exhibits a phase transition of topological nature. This model is a pure gauge theory with $\mathbb{Z}_2$ variables residing on the links (i.e., no matter fields). Its Hamiltonian is given by
\be
H = -J \sum_{\rm P} U_{\rm P} = -J \sum_{\rm P} \prod_{\ell \in P} S_{\ell}
\; ,
\label{eq1}
\ee
where $S_{\ell} = \pm 1$ are Ising variables placed on the links $\ell$ of the cubic lattice. The product over $\ell$ runs over all edges of a plaquette labeled by ${\rm P}$, and the 
sum over ${\rm P}$ runs over all plaquettes of the lattice~\footnote{On a cubic lattice of linear size $L$, there are ${\cal N}\, = 3\, L^3$ total number of plaquettes.}. Accordingly, each plaquette contributes through the product of its link variables, $U_{\rm P} = \prod_{\ell \in P} S_{\ell}$, which also takes values $\pm 1$. By analogy with spin-glass terminology, a plaquette is called simple or frustrated if $U_{\rm P} = +1$ or $U_{\rm P} = -1$, respectively. Here, $J>0$ denotes the coupling constant.

In dimension $D = 3$ and in thermal equilibrium, the model~\eqref{eq1} is Kramers--Wannier (KW) dual~\cite{10.1063/1.1665530} to the 3D ferromagnetic Ising model. The critical inverse temperature of the gauge model, $\beta_{\rm c}$, is related to that of the Ising model, $\beta_{\rm c}^*$, via the relation $2 \beta_{\rm c} J = - \ln \tanh (\beta^*_{\rm c} J^*)$, yielding the critical temperature $T_{\rm c} \simeq 1.313346\, J/k_{\rm B}$ (where $k_{\rm B}$ is the Boltzmann constant). In the absence of a local order parameter, the phases across the critical point are characterized in terms of global correlation functions, e.g., the Wilson loop~\cite{PhysRevD.10.2445,RevModPhys.51.659} and the Polyakov loop~\cite{POLYAKOV197582}, corresponding to a low-temperature deconfinement phase and a high-temperature confinement phase. The interpretation of different critical exponents in terms of these global observables remained challenging until recently~\cite{33q3-g68k}, when the model~\eqref{eq1} was investigated from a percolation perspective by defining various geometrical objects (see below). Intriguingly, the corresponding percolation temperature $T_{\rm p}$ coincides with $T_{\rm c}$, and the percolation exponents thus obtained provide direct access to the thermal critical exponents of the model~\eqref{eq1}. These exponents are also in strong agreement with those of the 3D Ising model~\cite{PhysRevLett.28.240,PhysRevE.97.043301}, supporting the emergence of the same universality class in the dual gauge model.

\begin{figure}[t!]
	\centering
	\rotatebox{0}{\resizebox{.60\textwidth}{!}{\includegraphics{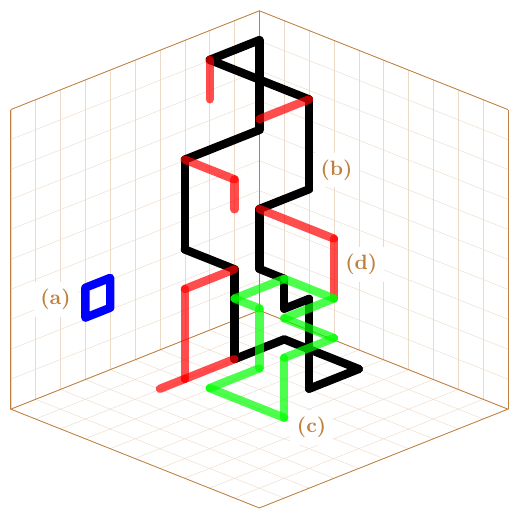}}}
\caption{Schematic of different geometrically defined objects in Wegner's three-dimensional gauge model: (a) an elementary loop formed by flipping a single spin with respect to the 
background configuration (in blue), (b) a system-spanning line loop (in black), (c) a short finite-sized loop (in green), and (d) dangling ends (in red) added to the loop in (b) with a temperature-dependent probability; see the text for further details. While (b) and (c) represent line loops formed by piercing frustrated plaquettes, 
the structure formed jointly by (b) and (d) constitutes a Fortuin--Kasteleyn (FK) cluster.}
	\label{fig1}
\end{figure}

However, the concept of universality is not limited to static exponents; dynamical classes~\cite{RevModPhys.49.435,tauber2014critical,RevModPhys.76.663} are also identified by the values of dynamic critical exponents governing the dynamics, both in classical~\cite{PhysRevE.108.064131,PhysRevE.110.034120} and quantum~\cite{PhysRevB.81.012303,PhysRevB.88.201110} settings. These exponents are strongly influenced by the conservation laws imposed dynamically (e.g., nonconserved or conserved dynamics), as well as by the nature of the dynamics itself (local or non-local). Near a continuous phase transition, the characteristic timescale $t_{\rm c}$ associated with the dynamics diverges as a power of the correlation length $\xi_{\rm c}$, i.e., $t_{\rm c} \propto \xi_{\rm c}^{z_{\rm c}}$, where $z_{\rm c}$ is the dynamical critical exponent describing the critical relaxation of thermal observables (e.g., energy density). In a finite system the divergence becomes $t_{\rm c} \propto L^{z_{\rm c}}$, where $L$ is the linear system size. Interestingly, despite the KW duality, numerical studies report markedly different estimates of the exponent $z_{\rm c}$ in the gauge and Ising models.

Concerning measurements in the gauge model, a recent study~\cite{PhysRevE.111.054107}, based on non-equilibrium finite-size scaling of the energy density, finds $z_{\rm c} \simeq 2.610(15)$, while in equilibrium settings a value $z_{\rm c} \simeq 2.55(6)$ is reported~\cite{PhysRevB.111.115129}. Another study~\cite{PhysRevB.97.024432}, analyzing soft quenches within the Kibble--Zurek scaling framework, yields a slightly larger value, $z_{\rm c} \simeq 2.70(3)$. These values are significantly larger than the exponent $z_{\rm c} \simeq 2.02$ associated with the 3D Ising universality class~\cite{10.1063/1.338572,Wang_1995}. This discrepancy underscores the need for a dedicated investigation of critical dynamics under gauge invariance. Notably, there is considerable interest in dynamical exponents in gauge-invariant systems due to their applications in condensed matter physics~\cite{PhysRevLett.66.3071,PhysRevB.58.2827,PhysRevLett.87.197003,PhysRevLett.88.137004,PhysRevLett.92.097004,dudka2008gauge}.

In the present study, we explore the critical quench dynamics by employing \textit{single-spin-flip} Monte Carlo simulations, which ensure local gauge invariance. We consider two quench protocols: (I) quenches to the critical point $T = T_{\rm c}$ from a high-temperature percolation phase at $T = 2 \, T_{\rm c}$, and (II) quenches to $T_{\rm c}$ from the ground state at zero temperature. We introduce a dynamical critical exponent $z_{\rm p}$ for the relaxation of percolation observables during the critical quenches. We note that in percolation problems~\cite{stauffer2018introduction,Essam_1980,SABERI20151}, the mass fraction of the largest cluster plays the role of an order parameter, analogous to the magnetization. 
The interesting question is whether the value of $z_{\rm p}$ differs from the exponent $z_{\rm c}$ governing energy relaxation. To address this, we perform a detailed finite-size scaling analysis of various percolation observables during the dynamics.

Another related issue is to understand how a gauge-invariant system locally organizes itself during quench dynamics. In systems with local order parameters, the time evolution following a critical quench is also critical on the scale of a growing lengthscale. However, in systems with global order parameters (e.g., the present one), the scenario is not yet fully understood and requires careful investigation. An intriguing question is whether a \textit{dynamical} lengthscale also emerges in such systems. We investigate these aspects from a geometric standpoint~\cite{PhysRevE.76.061116,Blanchard_2012,PhysRevE.94.062146,PhysRevE.105.034131}. We expect that, during the evolution, the gauge spins \textit{locally} build geometrical structures with \textit{critical} fractal properties, while maintaining local gauge invariance. Obviously, the growth kinetics of these structures can be interpreted depending on the representation of interest. Following Ref.~\cite{33q3-g68k}, we work with loop and random-cluster representations and define geometrical line loops (using two different construction rules, see Sec.~\ref{S2}) and Fortuin--Kasteleyn (FK) clusters during the time evolution after a critical quench. We quantitatively characterize their morphology using geometric measures.

The main findings of our study are as follows:
\begin{enumerate}[label=\arabic*., nosep]
\item
The critical relaxation of the percolation order parameter is governed by a dynamical exponent $z_{\rm p} \simeq 2.6$, consistent with the corresponding exponent for energy relaxation, $z_{\rm c}$.
\item
The value of $z_{\rm p}$ is robust, within error bars, with respect to both the quench protocol and the choice of geometrically defined objects.
\item
The time evolution of the number statistics $N(s,t)$ of geometrical objects of size $s$, following a quench from the percolation phase, supports a dynamic scaling framework governed by a time-dependent lengthscale $\xi_{\rm p}(t) \sim t^{1/z_{\rm p}}$.
\item
The statistics of geometrical objects exhibit rich crossovers between distinct dynamical regimes.
\end{enumerate}

We structure this paper as follows. In Sec.~\ref{S2} we provide details of the geometrically defined objects, while Sec.~\ref{S2_1} discusses the numerical implementation of the simulations. Secs.~\ref{S3} and \ref{S4} present our main results. In Sec.~\ref{S3}, we investigate the dynamical critical exponents, while Sec.~\ref{S4} analyzes the morphology of the geometrical objects during the time evolution. Finally, in Sec.~\ref{S5} we discuss the conclusions of this work. Appendix~\ref{collapse} discusses the procedure of extracting dynamical exponent from a data collapse.

\section{Geometrical objects}
\label{S2}

We define the geometrical objects of interest, namely geometric line loops and Fortuin--Kasteleyn (FK) clusters, and describe their construction.

\subsection{Geometrical line loops}

In the ground state of the gauge model~\eqref{eq1} at temperature $T=0$, all plaquettes are simple, $U_{\rm P}=+1$, and there are no plaquettes with $U_{\rm P}=-1$. For any nonzero $T$, however small, such frustrated plaquettes do exist. Therefore, frustrated plaquettes correspond to thermal excitations on top of the ground state. For the construction of geometric line loops, we assume that each frustrated plaquette is pierced perpendicularly by a line segment. Due to the discrete Bianchi identity, an even number of plaquettes is always frustrated in each elementary cube of the lattice configuration. Therefore, when the line segments are joined, they always form closed line loops, see Fig.~\ref{fig1}. Interestingly, these loops correspond to the low-temperature (LT) expansion graphs~\cite{33q3-g68k} of the partition function:
\be
\label{eq2}
{\cal Z} = \sum_{\text{config.}} \prod_{P} {\rm e}^{\beta J U_{\rm P}} \propto {\rm e}^{{\cal N} \beta J} \sum_{\ell} g_{\ell} \left( {\rm e}^{-2 \beta J} \right)^{\ell}, \quad \ell = 0, 4, 6, \ldots
\; ,
\ee
where $g_{\ell}$ is the degeneracy of the configurations with total loop length $\ell$, and ${\cal N}\, (=3\, L^3)$ is the total number of plaquettes. At low temperatures, thermal fluctuations generate short loops, while at high temperatures loops percolating across the system boundaries are formed.

In the literature, similar line-like objects have been identified in a wide range of systems, including cosmic strings in the early universe~\cite{Kibble_1976,PhysRevD.30.2036}, lines of darkness in light fields~\cite{PhysRevLett.100.053902}, vortex loops in the $XY$ model~\cite{PhysRevLett.57.1358,KAJANTIE2000114} and complex field theories~\cite{PhysRevB.72.094511,PhysRevE.94.062146,Kobayashi_2016}, as well as, more recently, disclination lines in active nematics~\cite{PhysRevLett.132.258301} and vortex strings in a holographic superfluid~\cite{xia2026evolutionvortexstringsthermal}. We also discuss a technical aspect of the line loops (also see Ref.~\cite{33q3-g68k}). The identification of different closed loops in a given spin configuration is an ambiguous task, as loops, especially those which are long, can have branches. These arise within individual cubic cells when more than two plaquettes (i.e., four, or six) are frustrated, e.g., see loops (b) and (c) in Fig.~\ref{fig1}.

There are mainly two methods proposed in the literature~\cite{KAJANTIE2000114,PhysRevB.72.094511,PhysRevE.94.062146,Kobayashi_2016} to resolve these branches, and we consider both in the present paper. One is the maximal reconnection method, where connections at a branching point are chosen so as to maximize the total length of the resulting loop. Another is the stochastic reconnection method, where connections at a branching point are chosen stochastically with a dice. In this method, loops are typically shorter than those obtained from the maximal reconnection method. Since there are fewer branches at $T \le T_{\rm c}$, the statistics of loops obtained from different methods remain the same at such temperatures. However, at higher temperatures $T > T_{\rm c}$, they show significant differences.

\subsection{Fortuin--Kasteleyn (FK) clusters}

FK clusters are obtained when a spin model is written in a random-cluster representation~\cite{FORTUIN1972536,Coniglio_1980}, where spin degrees of freedom are integrated out and \textit{new} bond variables taking binary values are introduced. These were originally proposed to understand phase transitions in the Ising/Potts spin models.

FK clusters were recently defined by us in Ref.~\cite{33q3-g68k} for the gauge model~\eqref{eq1}. In its random-cluster representation, the partition function is given by
\be
{\cal Z} = \sum_{\text{config.}} \prod_{\rm P} \sum_{n_{\rm P} = 0,1} {\rm e}^{\beta J} \left[ (1-q)\, \delta_{n_{\rm P},0} + q\, \delta_{U_{\rm P},1}\, \delta_{n_{\rm P},1} \right]
\; ,
\ee
where a bimodal random variable $n_{\rm P}=0,1$ is assigned to each plaquette of the cubic lattice: if $n_{\rm P} = 1$, the plaquette is occupied, and if $n_{\rm P} = 0$, it is empty. The variable $q$ sets a temperature-dependent probability rule:
\be
q = 1 - {\rm e}^{-2\beta J}
\; .
\label{prob}
\ee
In this representation, a simple plaquette is empty with probability $(1-q)$, while a frustrated one is always empty. An FK cluster is then constructed from neighboring empty plaquettes, see Fig.~\ref{fig1}. Note that two plaquettes are considered neighbors if they belong to the same cube. These clusters are not completely closed due to dangling ends (contributions from simple empty plaquettes).

During the dynamics at time $t$ after a quench to the critical point at $t=0$, we construct the critical FK clusters using $\beta = \beta_{\rm c}$ in Eq.~\eqref{prob}.

\section{Numerical Details}
\label{S2_1}

We investigate the critical dynamics of the gauge model~\eqref{eq1} on a cubic lattice with periodic boundary conditions (PBCs) in all three spatial directions. Since we consider two different quench protocols for the critical dynamics --- quenches to the critical point $T = T_{\rm c}$ from a high-temperature phase at $T = 2 \, T_{\rm c}$, and from the ground state at $T = 0$ --- we prepare the initial system configuration at time $t=0$ accordingly. For the $T=0$ initial state, all link spins are fixed to $+1$, which is one of the ground-state configurations. Since we work solely in the plaquette representation, it does not matter which ground-state configuration is chosen.

The link spins in Hamiltonian~\eqref{eq1} do not have intrinsic dynamics, and therefore we employ Markov chain Monte Carlo (MCMC) to evolve the system during the quench dynamics. We assume that the system is coupled to a heat bath, which induces stochastic spin flips according to the following master equation:
\begin{eqnarray}
	\nonumber
	\frac{d}{dt} P\left(S_1, \dots, S_{\ell}, \dots, S_N; t \right)
	&=& - \sum_{\ell}^{N}
	W_{S_{\ell} \rightarrow -S_{\ell}}
	P\left(S_1, \dots, S_{\ell}, \dots, S_N; t \right)
	\\
	&&
	+ \sum_{\ell}^{N}
	W_{-S_{\ell} \rightarrow S_{\ell}}
	P\left(S_1, \dots, -S_{\ell}, \dots, S_N; t \right)
	\; ,
	\label{master}
\end{eqnarray}
where $P\left(S_1, \dots , S_{\ell}, \dots , S_N; t \right)$ is the joint probability distribution of the link spins at time $t$, and $W$ is the transition rate~\cite{metropolis1949monte,newman1999monte}
\be
W(S_{\ell} \to -S_{\ell}) = \frac{1}{{\cal N}}\min \left\{1,e^{-\frac{\Delta E}{T}}\right \}
\; ,
\label{metrop}
\ee
which satisfies the detailed balance condition. Here, $\Delta E$ is the energy difference associated with a proposed link-spin flip, and ${\cal N}$ is the total number of link spins. The system size is ${\cal N} = 3L^3$, where $L$ is the linear size of the system. We note that a single link-spin flip changes the sign of all four plaquettes sharing that link, thereby ensuring that gauge invariance is preserved during the dynamics. The simulation time is measured in discrete Monte Carlo steps (MCS), each corresponding to ${\cal N}$ attempted link-spin updates.

For a robust numerical analysis, we consider cubic lattices of five different sizes: $L=12$, $L=16$, $L=24$, $L=32$, and $L=48$. For each system size, the numerical data are averaged over approximately $5 \times 10^4$ thermally uncorrelated runs initiated with different random seeds. To ensure statistical accuracy, the error bars for different observables are estimated using the Jackknife resampling technique~\cite{efron1982jackknife}. We consider relatively small system sizes ($L \le 48$), since the major computational effort is devoted to the efficient analysis of loop/cluster statistics from the initial stages of the dynamics up to equilibration timescales. We note that the construction and tracking of these objects are computationally demanding processes, with the computational cost increasing rapidly with $L$.

In the next section, we present our main results.

\section{Dynamical critical exponent}
\label{S3}

In this section, we calculate the dynamical critical exponents $z_{\rm c}$ and $z_{\rm p}$ associated with the relaxation of energy density and percolation observables, respectively.

\subsection{Relaxation of energy density}

We first explore the dynamical exponent $z_{\rm c}$ associated with the relaxation of the energy density, which has previously been estimated using different approaches~\cite{ben1990critical,PhysRevB.97.024432,PhysRevB.111.115129,PhysRevE.111.054107}. We note that, in the present study, the system is quenched to $T_{\rm c}$ from a high-temperature phase ($T = 2\,T_{\rm c}$), a setup that has not been considered in these earlier works. It is therefore useful to first benchmark the exponent $z_{\rm c}$ within the present setup.

Let us define the total energy density,
\bea
\label{Seq00}
e(L) &=& -\frac{J}{{\cal N}} \left\langle \sum_{\rm P} U_{\rm P} \right\rangle
\; ,\\ \nonumber
&=& J \left( 2 \frac{\langle {\cal N}_{\rm-} \rangle}{{\cal N}} - 1 \right)
\; ,
\eea
where the symbol $\langle \ldots \rangle$ denotes the thermal average, and $\langle {\cal N}_{\rm -} \rangle$ is the average number of frustrated plaquettes. 
Herein, the coupling constant $J$ is fixed to unity. In equilibrium at $T_{\rm c}$, the total energy $e(L)$ can be separated into regular and critical parts as
\be
e(L) = e_0 + a L^{-\Delta_{\rm \epsilon}},
\label{ee2}
\ee
where $e_0$ is the constant regular part and the second term $a L^{-\Delta_{\rm \epsilon}}$ represents the critical (singular) part, with $a$ being a constant and $\Delta_{\rm \epsilon} = D - 1/\nu$  
the scaling dimension of the energy operator~\cite{33q3-g68k}.

\begin{figure}[t!]
	\centering
	\rotatebox{0}{\resizebox{.48\textwidth}{!}{\includegraphics{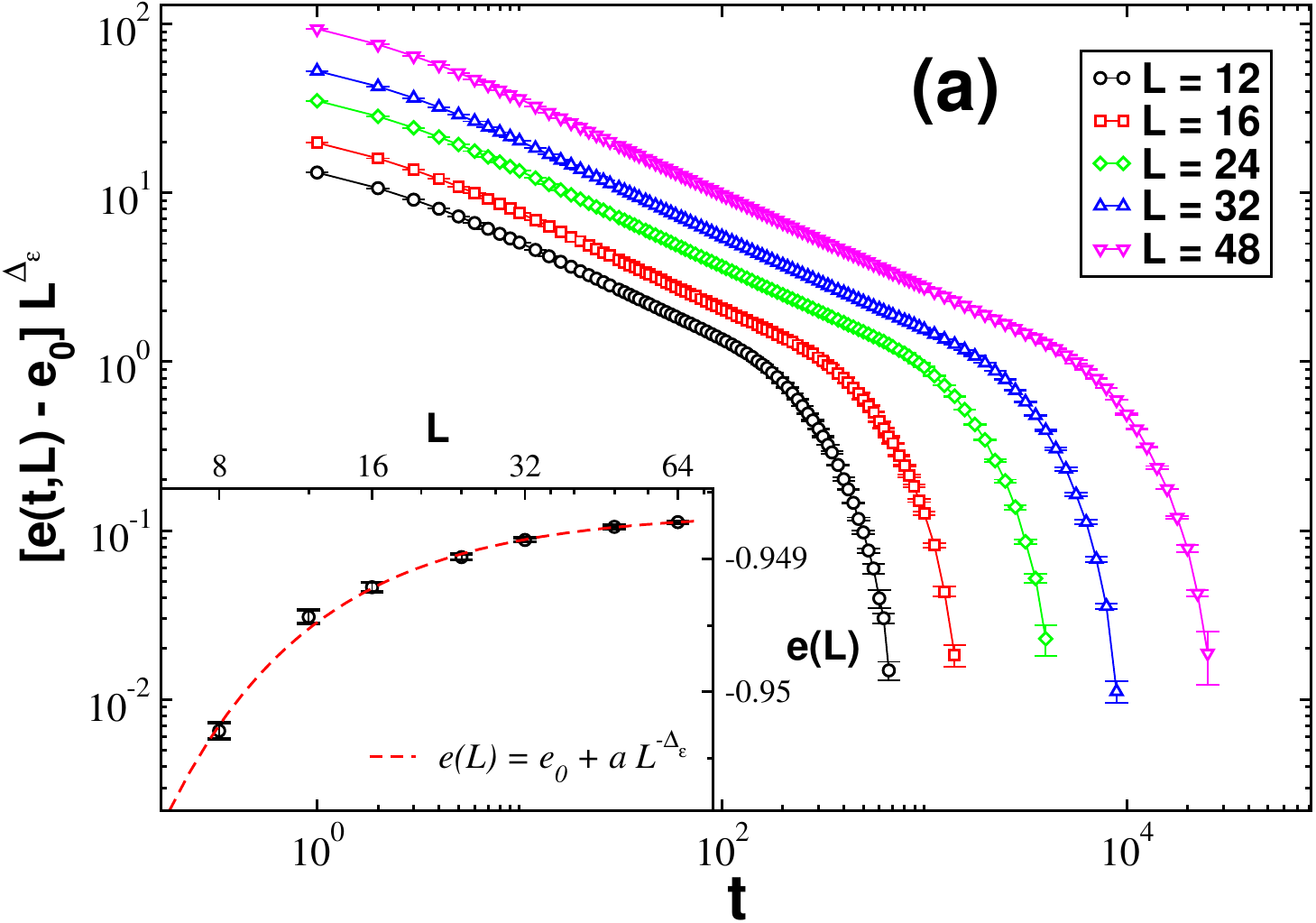}}}
	\rotatebox{0}{\resizebox{.48\textwidth}{!}{\includegraphics{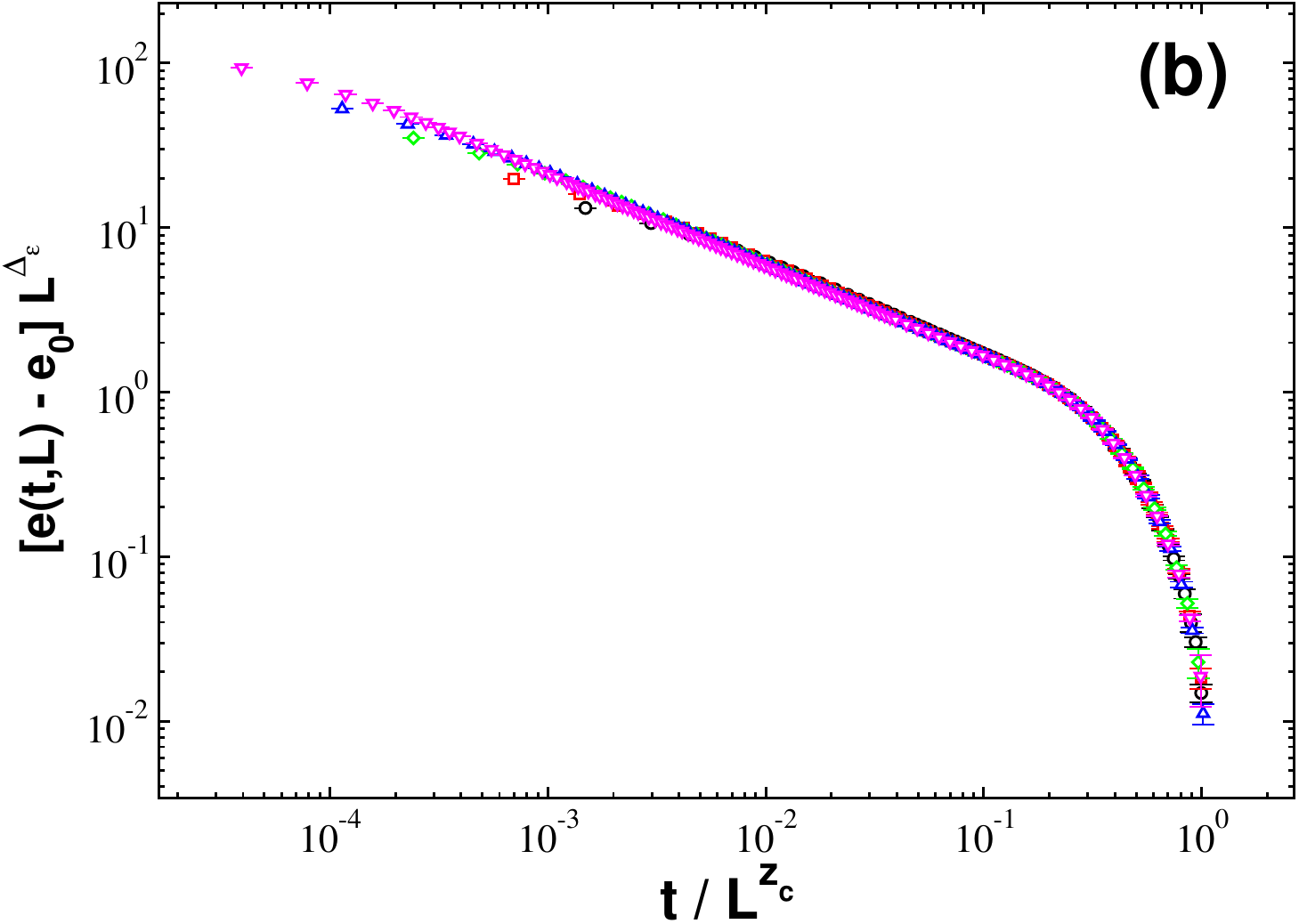}}}
	\caption{(a) Plot of the scaled variable $[e(t, L) - e_0] L^{\Delta_{\rm \epsilon}}$ against time $t$ on a log-log scale for different system sizes $L$ (given in the key of panel (a)).
		In the inset, the regular part $e_0 \simeq -0.94864(3)$  is extracted from the equilibrium total energy $e(L)$ at $T_{\rm c}$ by a fit of the form~\eqref{ee2} with $\Delta_{\rm \epsilon} \simeq 1.413$ and $a \simeq -0.029$.
		(b) Plot against scaled time $t/L^{z_{\rm c}}$ on a log-log scale, where 
		$\Delta_{\rm \epsilon} = 1.413$ and $z_{\rm c} = 2.62$.}
	\label{fig2}
\end{figure}

After the quench, the total energy density depends on time, $e \equiv e(t,L)$. For quenches to $T_{\rm c}$ from a high-$T$ state, we propose a \textit{simple} finite-size scaling Ansatz
\be
e(t, L) - e_0 \equiv f_e\!\left( \frac{t}{L^{z_{\rm c}}} \right ) L^{-\Delta_{\rm \epsilon}}
\; ,
\label{ee3}
\ee
where $e_0$ is the asymptotic (regular) energy density at $T_{\rm c}$, and $z_{\rm c}$ is the dynamical exponent governing the relaxation of the singular contribution. Note that the non-equilibrium scaling function $f_e$ depends only on a single scaling variable, $t/L^{z_{\rm c}}$. This differs from the scaling proposed in Ref.~\cite{PhysRevE.111.054107} for quenches to $T_{\rm c}$ from $T_{\rm c}^{+}$ (i.e., within the critical region), where the temperature deviation also enters as a relevant scaling variable. In the present case, since the system is quenched from a very high-$T$ ($T = 2\,T_{\rm c}$), any critical contribution at time $t = 0$ is negligible, and therefore $f_e$ depends only on $t/L^{z_{\rm c}}$. That said, after the quench to $T_{\rm c}$ the regular part also evolves dynamically (although with a much shorter relaxation time than the critical part), so deviations from the above scaling may appear at early times.

In Fig.~\ref{fig2}, the numerical results for the energy density are shown. First, the regular part of $e(L)$ in equilibrium at $T_{\rm c}$ is obtained; see the inset in panel (a). A fit to the form~\eqref{ee2} with $\Delta_{\rm \epsilon} = 1.413$~\cite{33q3-g68k} yields $e_0 \simeq -0.94864(3)$ within an acceptable range of $\chi^2$ values. Using this $e_0$, the scaled variable $[e(t, L) - e_0] L^{\Delta_{\rm \epsilon}}$ is plotted versus time $t$ in the main panel of (a) for different values of $L$. In (b), the scaling relation~\eqref{ee3} is tested by plotting the data in (a) 
against the scaled time $t/L^{z_{\rm c}}$. The exponent $z_{\rm c}$ is treated as a free parameter and is obtained by optimizing the scaling collapse (see procedure in Appendix~\ref{collapse}).
We find $z_{\rm c} \simeq 2.60(3)$, which is in agreement with recent estimates~\footnote{References~\cite{PhysRevE.111.054107} and \cite{PhysRevB.111.115129} report $z_{\rm c}\simeq2.610(15)$ and $z_{\rm c}\simeq2.55(6)$, differing from our estimate by only $0.3$ and $0.7$ combined standard deviations, respectively. The value $2.70(3)$ reported in Ref.~\cite{PhysRevB.97.024432} differs by about $2.4$ combined standard deviations. Since these studies employ different protocols (critical quench, equilibrium relaxation, and slow cooling), observables, and system sizes, the quoted statistical uncertainties do not reflect possible protocol-dependent corrections to scaling or other systematic effects.}. The obtained value for $z_{\rm c}$ is clearly different from the one for 3D Ising model, $z_{\rm c} \simeq 2.02$~\cite{10.1063/1.338572,Wang_1995}.

One can also analyze the plaquette densities associated with different geometrical objects. For example, the plaquette density of the line loops, defined as the fraction of frustrated plaquettes, is given by $e_{l} = \langle {\cal N}_{\rm -} \rangle / {\cal N}$. Similarly, the plaquette density of FK clusters can be investigated. It is related to $e_{l}$ as
\be
e_{\rm FK} = (1-q_{\rm c}) (1-e_{l}) + e_{l}
\; ,
\label{ep}
\ee
where $q_{\rm c} = 1 - {\rm e}^{-2\beta_{\rm c} J}$. Notice that in the random-cluster representation all frustrated plaquettes are always empty, while simple plaquettes are empty with probability $1-q_{\rm c}$. Since both $e_{l}$ and $e_{\rm FK}$ are directly proportional to the energy density~\eqref{Seq00} at a given time, they yield the same dynamical exponent $z_{\rm c}$ as above. For brevity, we do not show these data here. We also remark that both loops and FK clusters share a common exponent $\nu_{\rm p}$, which is identical to the thermal exponent of the energy operator, i.e., $\nu_{\rm p} = \nu$~\cite{33q3-g68k}.

\begin{figure}[t!]
	\centering
	\rotatebox{0}{\resizebox{.85\textwidth}{!}{\includegraphics{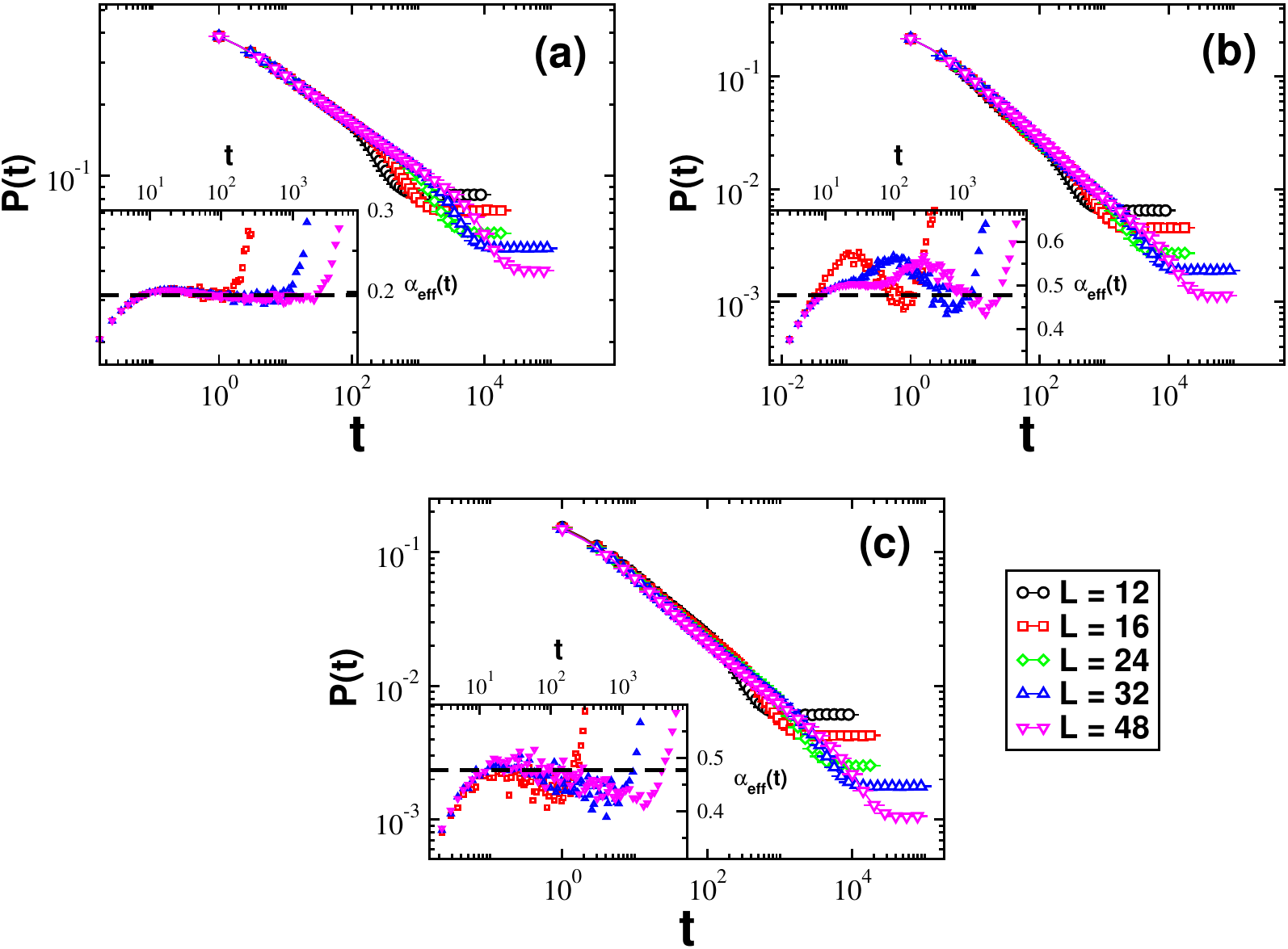}}}
	\caption{Percolation strength $P(t)$ vs. time $t$ on a log-log scale for (a) FK clusters, (b) line loops with maximal connection, and (c) line loops with stochastic connection, in systems of different linear sizes $L$ (see the key) quenched from $T= 2\, T_{\rm c}$ to $T_{\rm c}$. In the insets, the effective exponent $\alpha_{\rm eff}(t)$ is plotted against $t$ for $L=16,\, 32,\, 48$. 
	The horizontal dashed line in each inset represents the exponent $\beta_{\rm p}/(\nu_{\rm p} z_{\rm p})$; see the main text for details.}
	\label{fig3}
\end{figure}

\begin{figure}[t!]
	\centering
	\rotatebox{0}{\resizebox{.85\textwidth}{!}{\includegraphics{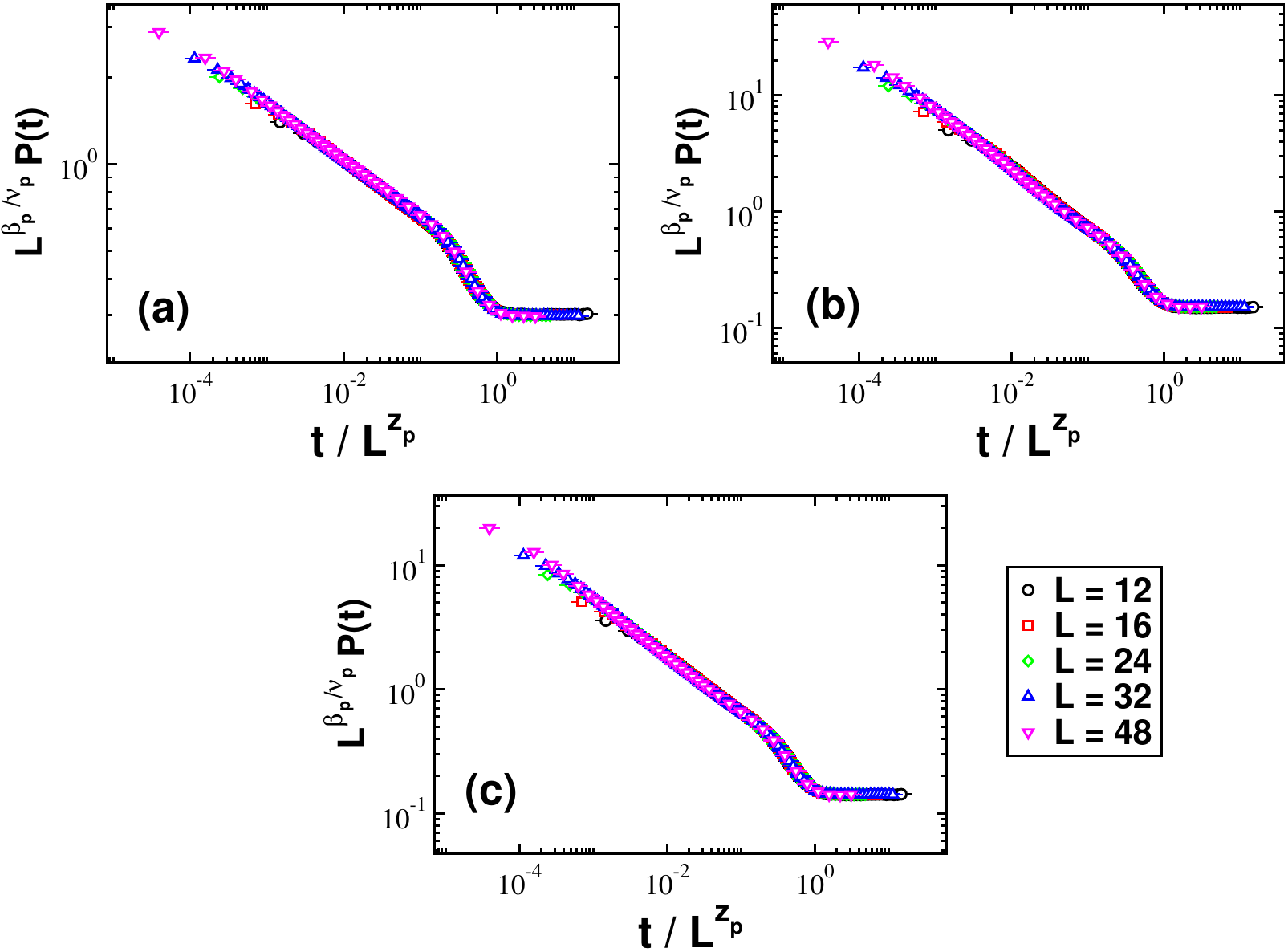}}}
	\caption{Scaled percolation strength $L^{\beta_{\rm p}/\nu_{\rm p}} P(t)$ vs. scaled time $t/L^{z_{\rm p}}$ on a log-log scale for (a) FK clusters, 
	(b) line loops with maximal connection, and (c) line loops built with the stochastic rule, in systems of different linear sizes $L$ (see the key) quenched from  
	$T= 2\, T_{\rm c}$ to $T_{\rm c}$.}
	\label{fig31}
\end{figure}

\subsection{Relaxation of percolation observables}

We now analyze various percolation observables to determine the corresponding dynamical exponent $z_{\rm p}$ and to investigate whether this exponent is related to the exponent $z_{\rm c}$ associated with the relaxation of the energy density.

For this purpose, we first consider the percolation strength $P$, a quantity that acts as an order parameter in percolation theory. It is defined as the thermal average of the mass fraction $m$ of the largest geometrical object in a given configuration,
\be
\label{P}
P = \langle m \rangle
\; , \; \; \; \; \qquad\quad m = s_l/(3L^3)
\; ,
\ee
where $s_l$ is the number of plaquettes contained in that object. We will see how this quantity behaves during the quench to the percolation temperature (i.e., $T_{\rm c}$). Following the \textit{short-time critical dynamics} (STCD) approach~\cite{janssen1989new,doi:10.1142/S021797929800288X,Albano_2011,PhysRevE.108.064131}, we assume the following scaling form for the $k^{\rm th}$ moment of the mass fraction for a start from the percolating high temperature state,
\be
\label{scale_mk}
\langle m^k(t) \rangle = \ell^{-k \beta_{\rm p}/\nu_{\rm p}} M_k(\ell^{-z_{\rm p}}t,\ell^{1/\nu_{\rm p}}{\cal T},\ell^{-1}L)
\; ,
\ee
where $\ell$ is an arbitrary spatial rescaling parameter, ${\cal T}$ is the reduced temperature, $L$ is the system size, $\beta_{\rm p}$ and $\nu_{\rm p}$ are the percolation critical exponents~\cite{33q3-g68k}, and $z_{\rm p}$ is the dynamical exponent describing the relaxation of the moment $\langle m^k(t) \rangle$. We note that the subscript ${\rm p}$ (here and throughout the paper) refers to the percolation origin.

\begin{figure}[t!]
	\centering
	\rotatebox{0}{\resizebox{.85\textwidth}{!}{\includegraphics{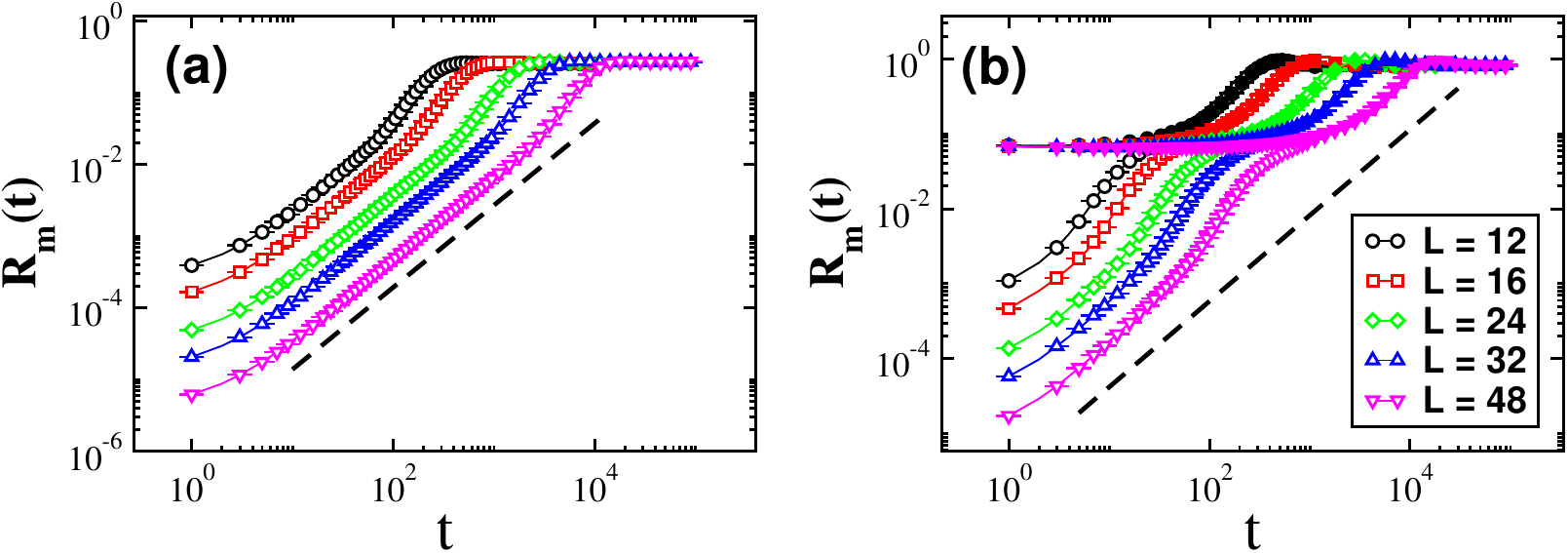}}}
	\rotatebox{0}{\resizebox{.85\textwidth}{!}{\includegraphics{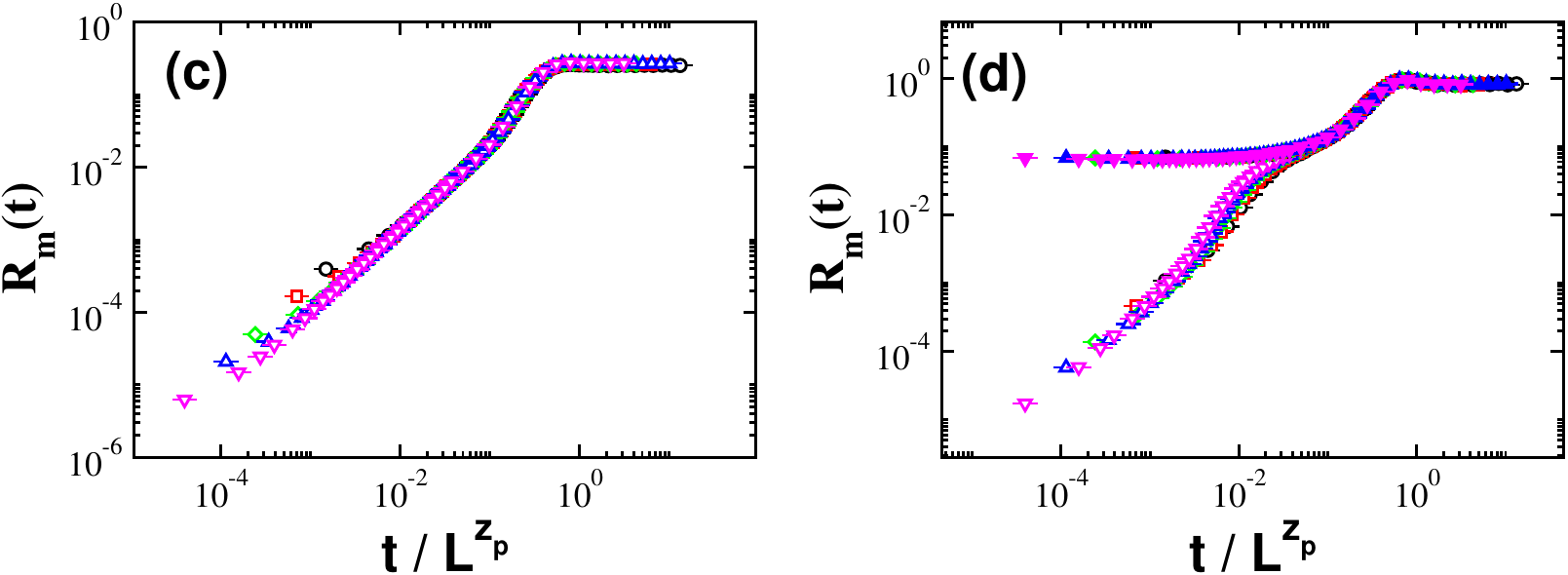}}}
	\caption{Relative variance $R_{\rm m}(t)$ vs. time $t$ on a log-log scale for (a) FK clusters, and (b) line loops, in a system of different linear sizes $L$ (given in the key in panel (b)) 
	quenched from $T= 2\, T_{\rm c}$ to $T_{\rm c}$. Panels (c) and (d) plot data in (a) and (b) against the rescaled time $t/L^{z_{\rm p}}$, respectively.  
	In (b) and (d) the datasets with empty symbols represent the loops constructed with the maximal rule, while those with filled ones represent the loops built with the stochastic rule. 
	The dashed lines in panels (a)--(b) indicate the law $R_{\rm m}(t) \sim t^{D/z_{\rm p}}$, with $z_{\rm p} = 2.62$ and $D=3$.}
	\label{fig5}
\end{figure}

During the critical dynamics (${\cal T}=0$), time and the spatial rescaling parameter are related through dynamic scale invariance. Choosing the rescaling factor as $\ell\sim t^{1/z_{\rm p}}$ therefore yields
\be
\label{m12}
P(t) \sim t^{- \beta_{\rm p}/(\nu_{\rm p} z_{\rm p})}
\; ,
\ee
which is valid in the regime $1 \ll \ell \ll L$. At this stage, $\ell$ is merely the renormalization-group rescaling parameter and should not be interpreted as a physically meaningful growing length. This question is examined independently in Section~\ref{S4} through the time-dependent cutoff and scaling collapse of the object-number distributions.

In Fig.~\ref{fig3}, we plot $P(t)$ versus time $t$ for different geometrical objects (see panels (a)--(c)) and different system sizes $L$, for a critical quench from a high-$T$ percolation phase. Let us first discuss panel (a) for the FK clusters. On the log-log scale, $P(t)$ decays linearly with $t$, indicating a power-law behavior. At late times, the decay becomes abrupt before reaching a saturation plateau, which signals equilibration. For smaller $L$, this plateau is reached at earlier times and at larger values of $P$. Different line-loop definitions in panels (b) and (c) exhibit similar behavior.

For a more sensitive test of the \textit{putative} power law $P(t) \sim t^{-\alpha_{\rm eff}}$, we plot in the insets the effective exponent
\be
\label{m13}
\alpha_{\rm eff} = - \frac{d \ln P(t)}{d \ln t}
\; ,
\ee
for the datasets with $L = 16,\,32,\,48$ shown in the main panels. If Eq.~\eqref{m12} holds, $\alpha_{\rm eff} \to \beta_{\rm p}/(\nu_{\rm p} z_{\rm p})$ asymptotically. Following Ref.~\cite{33q3-g68k}, the critical ratios are taken as $\beta_{\rm p}/\nu_{\rm p} \simeq 0.518$ and $\beta_{\rm p}/\nu_{\rm p} \simeq 1.265$ for FK clusters and line loops, respectively. 

For FK clusters, $\alpha_{\rm eff}$ initially increases (up to $t \lesssim 10$), then reaches a plateau around $0.197$ (with $z_{\rm p} \simeq 2.62$), before increasing again at late times due to equilibration effects. The time window over which $\alpha_{\rm eff}$ remains approximately constant also grows with $L$. However, a nontrivial behavior is observed for line loops, see the insets in (b)--(c). In (b), for loops constructed using the maximal rule, $\alpha_{\rm eff}$ oscillates around a value slightly above the exponent $\beta_{\rm p}/(\nu_{\rm p} z_{\rm p}) \simeq 0.48$ (i.e., $z_{\rm p} \simeq 2.62$), while in (c), for loops constructed using the stochastic rule, it oscillates around a slightly smaller value. Nevertheless, in both cases $\alpha_{\rm eff}$ initially approaches the expected exponent (shown by the horizontal dashed line) and remains close to it for longer times at larger system sizes (see data for $L=48$). Therefore, this deviation may be attributed to finite-size effects. We will return to this issue later.

\begin{figure}[t!]
	\centering
	\rotatebox{0}{\resizebox{.85\textwidth}{!}{\includegraphics{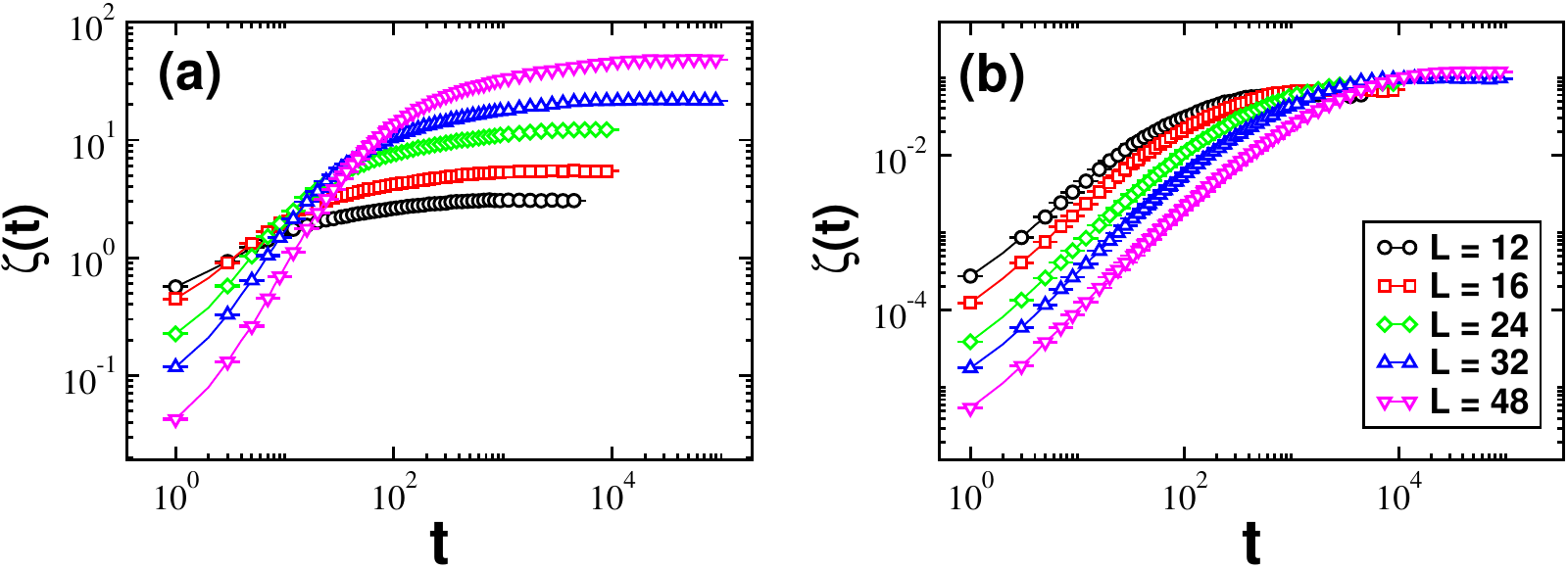}}}
	\rotatebox{0}{\resizebox{.85\textwidth}{!}{\includegraphics{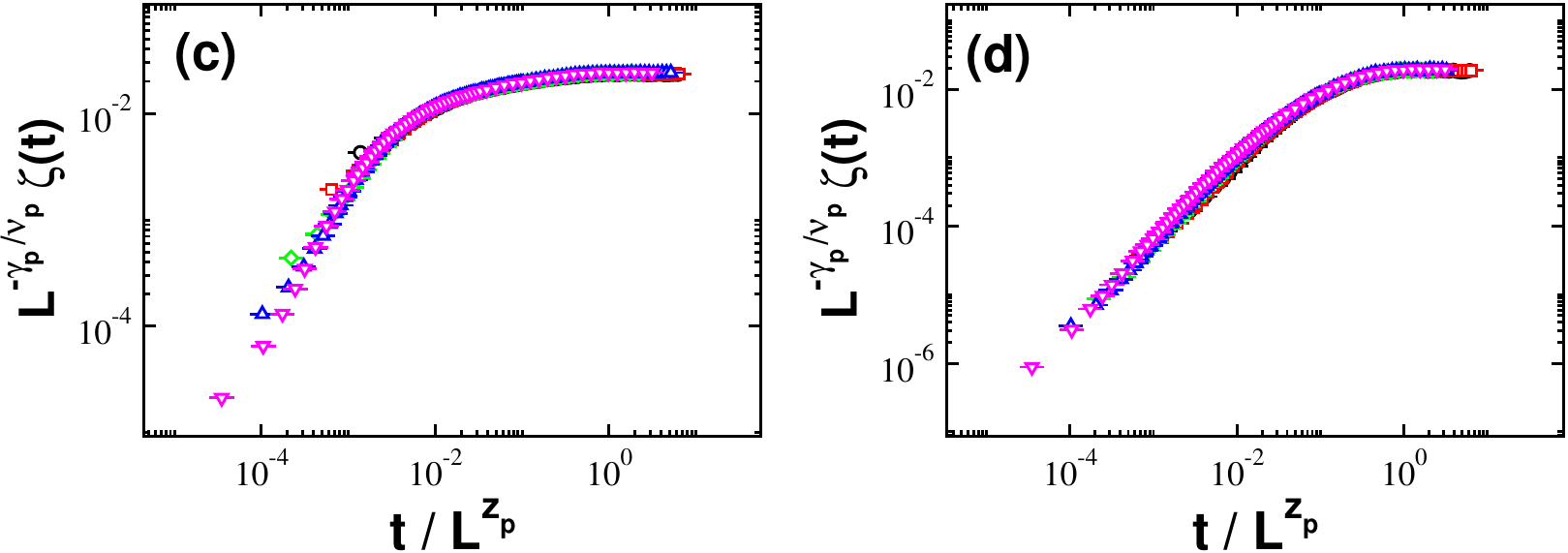}}}
	\caption{Plot of fluctuations $\zeta(t)$ vs. time $t$ for (a) FK clusters and (b) line loops of maximal rule, in systems with different linear sizes $L$ (given in the  key in (b)) 
	quenched from $T=0$ to $T_{\rm c}$. Panels (c) and (d) plot the data in (a) and (b) times $L^{-\gamma_{\rm p}/\nu_{\rm p}}$ against the rescaled time $t/L^{z_{\rm p}}$.}
	\label{fig6}
\end{figure}

Furthermore, we inspect the relaxation of $P(t)$ by setting $\ell = L$ and ${\cal T} = 0$ in the expression~\eqref{scale_mk}. This gives the following finite size scaling form,
\be
\label{m11}
P(t) \sim \frac{1}{L^{\beta_{\rm p}/\nu_{\rm p}}} f_P \left(\frac{t}{L^{ z_{\rm p}}} \right)
\; .
\ee
In Figs.~\ref{fig31}(a)--(c), the scaling form~\eqref{m11} is tested by plotting the scaled variable $L^{\beta_{\rm p}/\nu_{\rm p}} P(t)$ against scaled time $t/L^{\rm z_{\rm p}}$ for data in the main frame of Figs.~\ref{fig3}(a)--(c). By enabling a best scaling collapse in panels (a)--(c) (following procedure of Appendix~\ref{collapse}), we obtain a value of $z_{\rm p} \simeq 2.62(3)$ for FK clusters, and $z_{\rm p} \simeq 2.61(4)$ for both line loops, which are consistent with exponent $z_{\rm c}$ within statistical error bars.

Another quantity of interest is the relative variance of the mass fraction $m$,
\be
R_{\rm m}(t) = \frac{ \langle m^2(t)\rangle - \langle m(t) \rangle^2}{\langle m(t) \rangle^2}
\; ,
\ee
whose finite size scaling form can also be obtained from the relation~\eqref{scale_mk} as
\be
\label{m3}
R_{\rm m}(t) \sim f_R \left(\frac{t}{L^{ z_{\rm p}}} \right)
\; .
\ee
In panels (a)--(b) of Fig.~\ref{fig5}, the quantity $R_{\rm m}(t)$ is plotted versus $t$ for different system sizes $L$. In (a), for FK clusters, $R_{\rm m}(t)$ increases as a power law in time, $R_{\rm m}(t) \sim t^{3/z_{\rm p}}$~\cite{Albano_2011,PhysRevE.108.064131} with $z_{\rm p} \simeq 2.62$, and eventually saturates to an $L$-independent  value. A similar saturation is observed for line loops in panel (b). However, for line loops, $R_{\rm m}(t)$ does not exhibit a simple power-law growth. We note that a similar nontrivial behavior for line loops was observed in Figs.~\ref{fig3}(b)--(c) for $P(t)$. In panels (c)--(d) of the same figure, $R_{\rm m}(t)$ is plotted against the rescaled time $t/L^{z_{\rm p}}$ to test the scaling relation~\eqref{m3}. The best scaling collapse is obtained with $z_{\rm p} \simeq 2.61(3)$ for FK clusters and $z_{\rm p} \simeq 2.60(5)$ for both line loops.

The STCD is also useful for critical quenches from the $m=0$ state, i.e., a quench from the zero-temperature ground state to $T_{\rm c}$. Specifically, the fluctuations in $m$,
\be
\zeta(t) = L^D \left( \langle m^2(t)\rangle - \langle m(t) \rangle^2 \right)
\; ,
\label{z}
\ee
show the following finite size scaling form~\cite{Albano_2011},
\be
\zeta(t) = L^{\gamma_{\rm p}/\nu_{\rm p}} f_{\zeta} \left(\frac{t}{L^{ z_{\rm p}}} \right)
\; .
\label{z1}
\ee
In panels (a) and (b) of Fig.~\ref{fig6}, the quantity $\zeta(t)$ is plotted versus $t$ for FK clusters and loops constructed using the maximal rule, for a critical quench from $T=0$. For brevity, we do not show data for loops with the stochastic rule, as they behave identically to the maximal-rule loops below (and at) $T_{\rm c}$. For both objects, $\zeta(t)$ grows with time and eventually saturates to an 
$L$-dependent value. In panels (c)--(d) of the same figure, the scaling relation~\eqref{z1} is tested to estimate the exponent $z_{\rm p}$. From Ref.~\cite{33q3-g68k}, we use $\gamma_{\rm p}/\nu_{\rm p} \simeq 1.964$ for FK clusters and $\gamma_{\rm p}/\nu_{\rm p} \simeq 0.47$ for line loops, respectively. Using these values, the best scaling collapse is obtained for $z_{\rm p} \simeq 2.60(8)$ for FK clusters and  $z_{\rm p} \simeq 2.60(15)$ for line loops.

\begin{table}[t!]
	\centering
	\caption{Estimated values of the dynamical critical exponent $z_{\rm p}$ obtained from the optimization of scaling collapse for different geometrical objects and observables. The first two rows correspond to critical quenches from the high-temperature $T = 2 \, T_{\rm c}$, while the last row corresponds to a critical quench from $T=0$. Error bars are estimated using the goodness criterion $1 \lesssim S \lesssim 2.5$. The uncertainty propagated from the input static exponents is negligible compared to the quoted uncertainty of the dynamical exponents. For comparison, the dynamical exponent obtained from the relaxation of the energy density is $z_{\rm c}\simeq 2.60(3)$. See the main text and Appendix~\ref{collapse} for details.}
	\vspace*{0.2cm}
	\begin{tabular}{|l|c|c|c|}
		\hline
		Observable & FK clusters & Maximal-rule loops & Stochastic-rule loops \\
		\hline
		\hline
		$P(t)$          & 2.62(3) & 2.61(3)  & 2.60(3) \\
		$R_{\rm m}(t)$  & 2.61(3) & 2.61(4)  & 2.60(5) \\
		$\zeta(t)$      & 2.60(8) & 2.60(15) & 2.60(15) \\
		\hline
	\end{tabular}
	\label{table1}
\end{table}

We finally conclude that the relaxation exponent $z_{\rm p}$ of percolation origin is in good agreement with that of the energy density, $z_{\rm c}$, irrespective of the quench protocol and the type of geometrically defined object (see Tab.~\ref{table1}). The STCD-inspired relation~\eqref{scale_mk}, which is based on spatial renormalization via a rescaling parameter $\ell$, is robustly valid when $\ell = L$, i.e., when $L$ is the only relevant lengthscale. However, when $\ell$ is chosen to follow a time-dependent power-law form $\ell \sim t^{1/z_{\rm p}} \ll L$, the data for the 
FK clusters support this relation, while it appears inconclusive for line loops. We will examine this issue from a geometrical perspective in the next section~\ref{S4}, where the geometrical properties of the different objects are discussed.

\section{Morphology of geometrical objects}
\label{S4}

\begin{figure}[t!]
	\centering
	\rotatebox{0}{\resizebox{.48\textwidth}{!}{\includegraphics{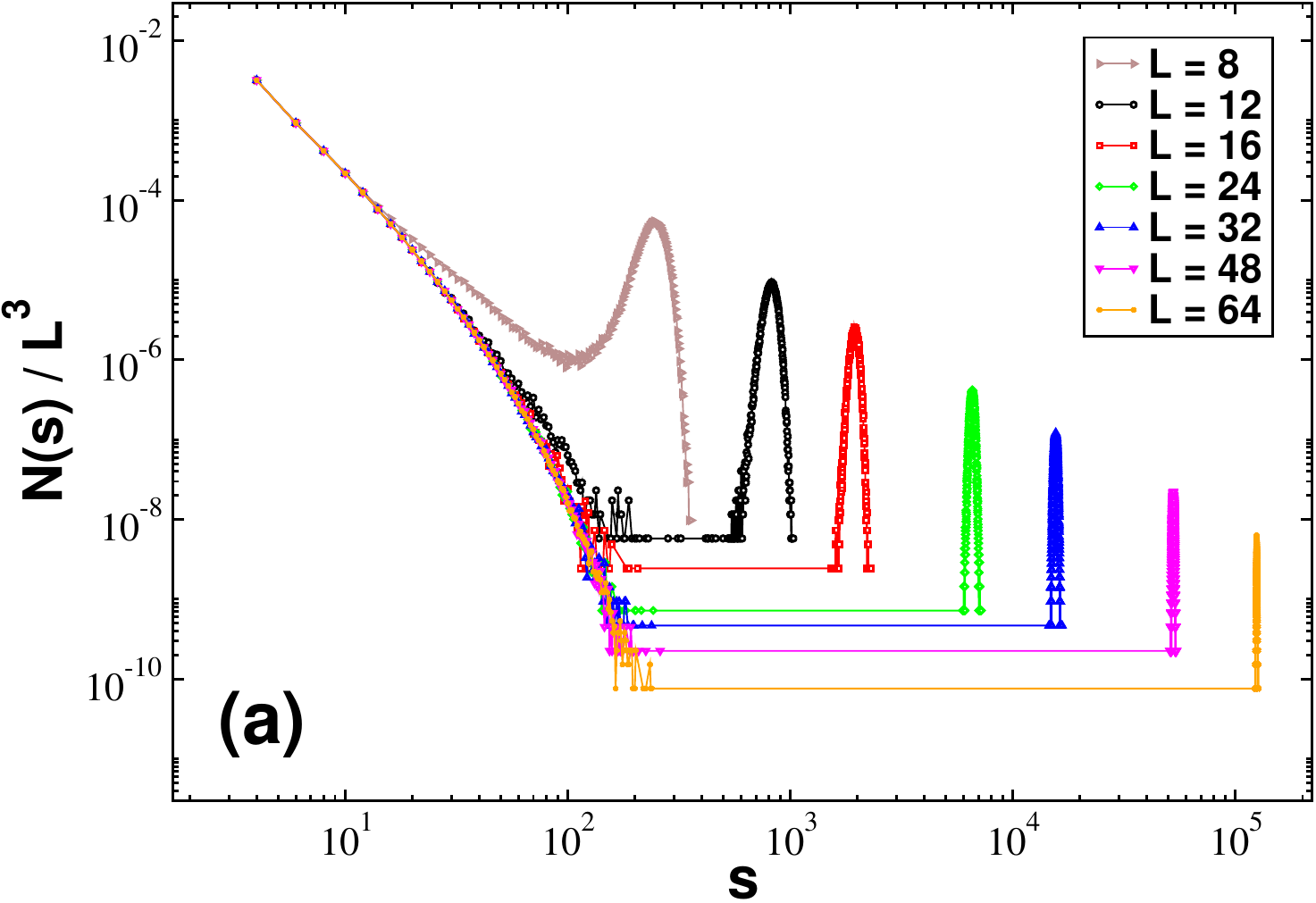}}}
	\rotatebox{0}{\resizebox{.48\textwidth}{!}{\includegraphics{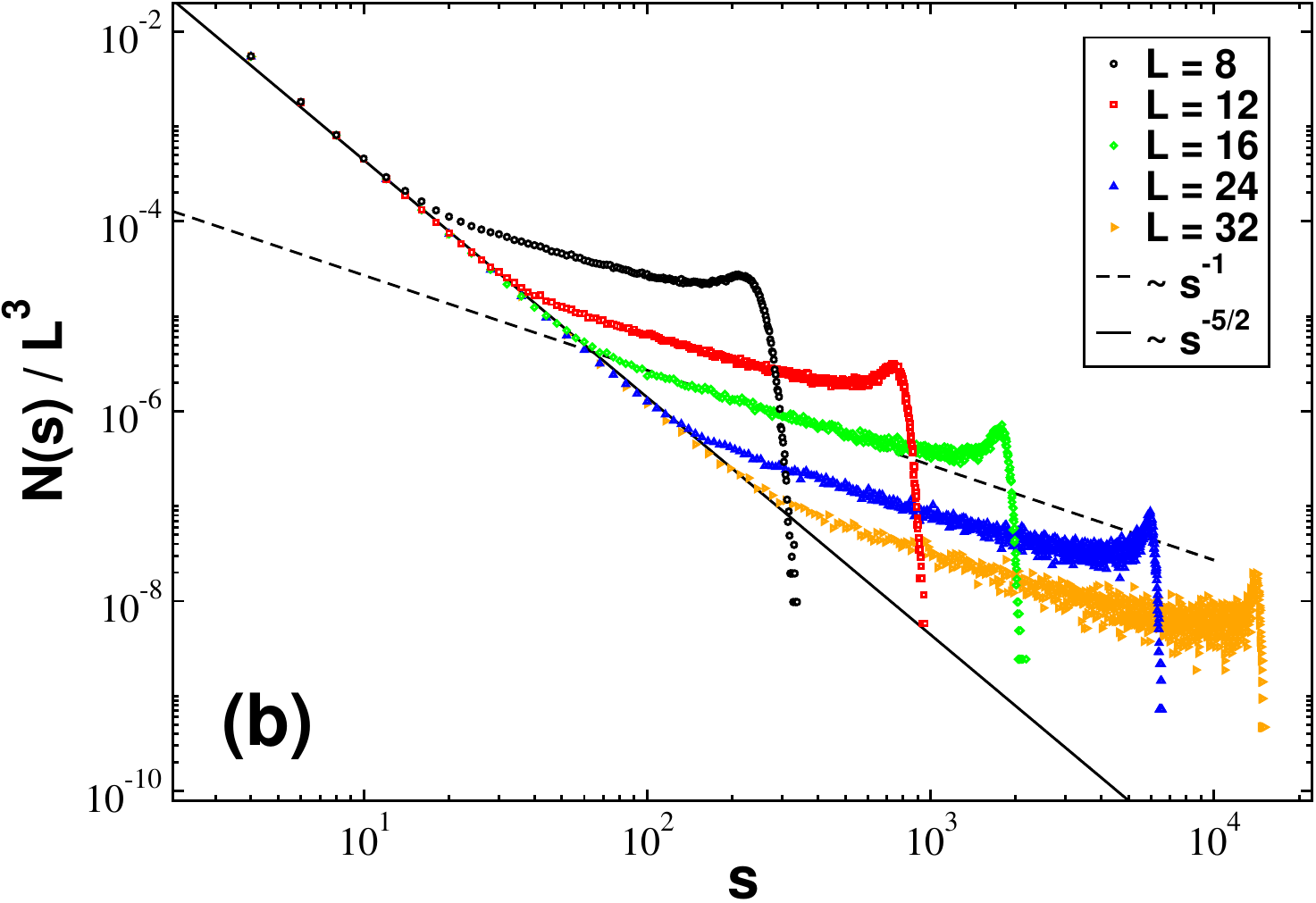}}}
	\rotatebox{0}{\resizebox{.48\textwidth}{!}{\includegraphics{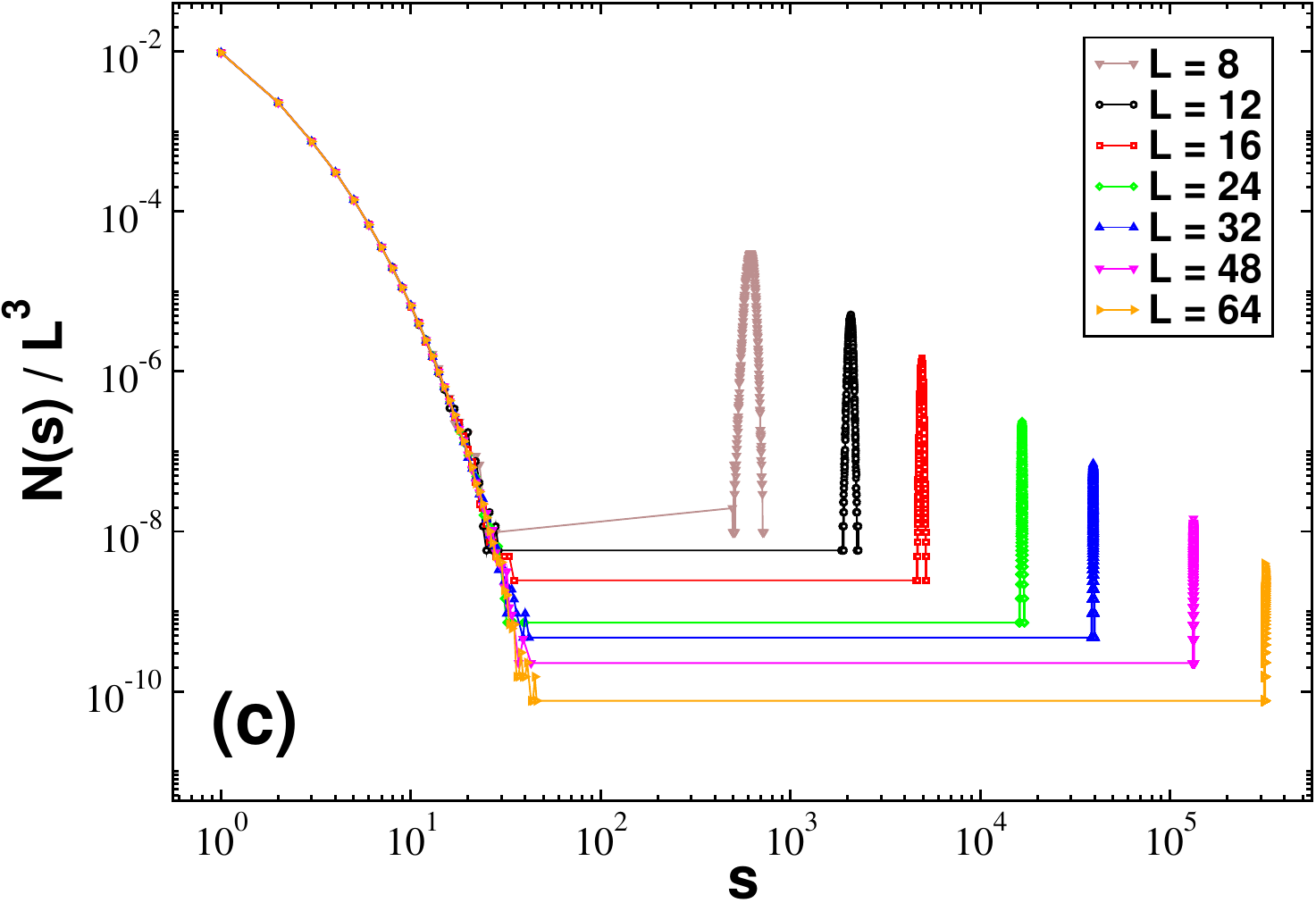}}}
	\caption{Number density per volume, $N(s)/L^3$, against size $s$, on a log-log scale, of (a) line loops constructed with the maximal rule, (b) line loops built with the stochastic rule, and (c) FK clusters, in equilibrium configurations of the gauge model with different linear sizes (given in the keys) at temperature  $T = 1.2~ T_{\rm c}$ (also see Supplemental Material in Ref.~\cite{33q3-g68k}). The lines joining different symbols in each dataset are shown as a guide to eye.}
	\label{fig61}
\end{figure}

In this section, we explore the morphology of different geometrical objects during the time evolution after a quench from the 
high temperature percolation phase to the critical point. For this purpose we probe the statistics of loops and FK clusters by means of their number profile --- the number $N(s)$ of such objects with mass (size) $s$. In equilibrium at temperatures below and close to $T_{\rm c}$, the asymptotic form of $N(s)$ normalized by volume $L^D$, i.e., the number density is given as
\be
N(s)/L^D \simeq \frac{{\rm e}^{-s \epsilon}}{s^{\tau_{\rm p}}}
\; ,
~~~~\qquad\quad
\epsilon \propto ( T_{\rm c} -T ) ^{1/\sigma_{\rm p}}
\; ,
\label{Seq3}
\ee
where $\tau_{\rm p} = 1 + D/D_{\rm f}$ and $\sigma_{\rm p} = 1/(D_{\rm f} \nu_{\rm p})$ are the critical exponents corresponding to the fractal dimension $D_{\rm f}$ and length tension $\epsilon$, respectively. In the limit $\epsilon \rightarrow 0$ at $T = T_{\rm c}$, the quantity $N(s)/L^{D}$ falls in a power law as
\be
N(s)/L^D \propto \frac{1}{s^{\tau_{\rm p}}}
\label{power_law}
\ee
in equilibrium. It is found in Ref.~\cite{33q3-g68k} that for FK clusters $\tau_{\rm p} \simeq 2.209$, while it is $\tau_{\rm p} \simeq 2.729$ for geometric line loops.

In our simulations, we obtain the time-dependent quantity $N(s,t)$ by averaging over the number of objects in various independent configurations of a fixed system size $L$ at time $t$ after the quench.

\subsection{Statistics in the percolation phase}

\begin{figure}[t!]
	\centering
	\rotatebox{0}{\resizebox{.99\textwidth}{!}{\includegraphics{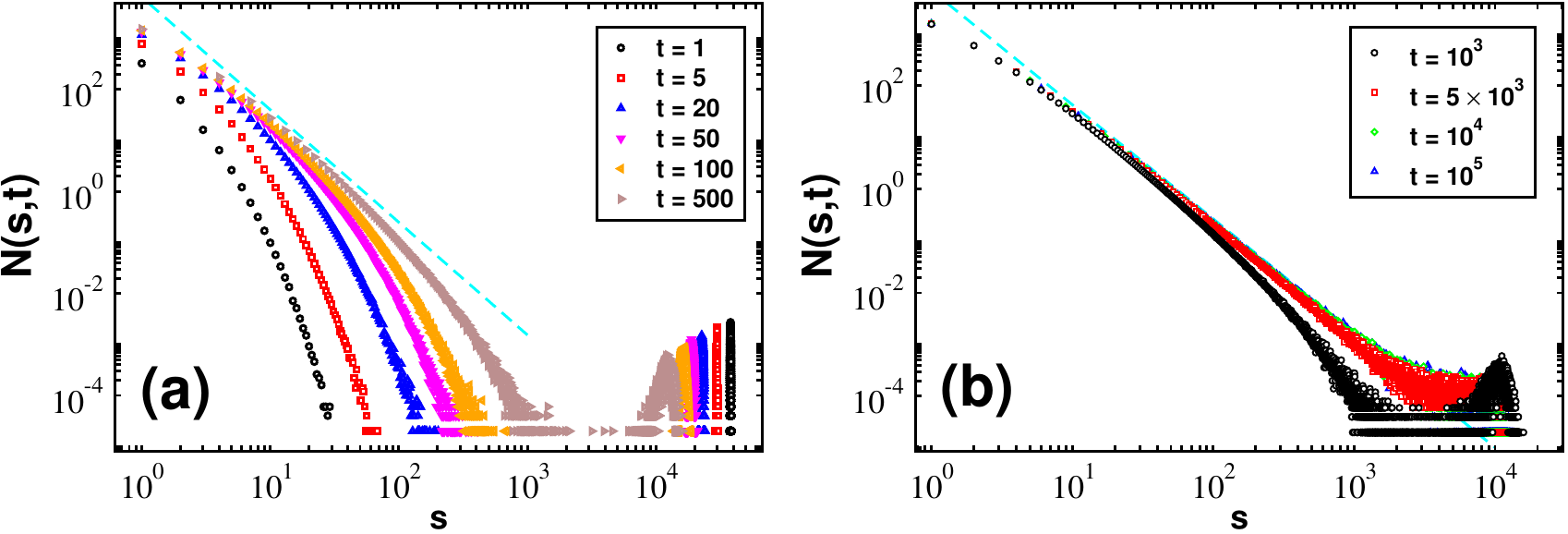}}}
	\caption{Number $N(s,t)$ vs. size $s$ (on a log-log scale) of FK clusters in a system of linear size $L=32$ quenched from $T= 2\, T_{\rm c}$ to $T_{\rm c}$. Different panels (a)--(b) show data in different time regimes (see the keys) subsequent to a quench at time $t=0$. The dashed line in both panels represent the power-law $N(s) \sim s^{-\tau_{\rm p}}$ with $\tau_{\rm p} = 2.209$.}
	\label{fig73}
\end{figure}

In this section, we first discuss the statistics of the quantity $N(s)$ for different geometrical objects in equilibrium in the high-temperature percolation phase (see also Ref.~\cite{33q3-g68k}).

In Fig.~\ref{fig61}, the quantity $N(s)/L^3$ of different geometrical objects is plotted against size $s$ for different system sizes $L$ at a temperature $T = 1.2 ~T_{\rm c}$. In panel (a), the data is shown for loops constructed with maximal rule. The number density of these loops shows an exponential fall at small $s$ and a huge bump at large value of $s$. Moreover, the bump position and the regime of initial exponential fall increases with $L$. This confirms the expected behavior of such loops --- the configuration at $T>T_{\rm c}$ is engulfed by a few very large loops and the remaining ones are very small both in size and number.

Notice that upon further increasing the temperature at fixed $L$, the large loops become even more gigantic at the cost of smaller ones, resulting in a further shift of the bump position and a reduction of the exponential regime~\cite{33q3-g68k}. Similar features have also been observed in other systems where line-like objects are constructed using the same rule, e.g., vortex tangles in $\mathcal{U}(1)$ complex field theories~\cite{PhysRevE.94.062146,Kobayashi_2016}. The FK clusters in panel (c) also exhibit analogous behavior.

In panel (b) of Fig.~\ref{fig61}, the data for line loops constructed using the stochastic method are shown, whose behavior is rather distinct. For small $s$, $N(s)/L^3$ follows a power law, $N(s)/L^3 \sim s^{-5/2}$, while for large $s$ it crosses over to a different power law, $N(s)/L^3 \sim s^{-1}$. Furthermore, the range over which both power laws hold increases with $L$. The small-scale behavior is consistent with Gaussian random walks~\cite{de1979scaling}, while the large-scale behavior is reminiscent of fully packed loop models~\cite{PhysRevD.49.4089,PhysRevLett.107.177202,PhysRevLett.111.100601} and emerges only due to long non-contractible loops. Such a two-fold behavior was analytically shown to exist in a simple model of strings on a periodic lattice~\cite{PhysRevD.49.4089}. To summarize,
\bea
N(s)/L^3 &\simeq& 
\left\{
\begin{array}{l}
	s^{-5/2}
	\; ,
	~~~~~~~~ s \ll L^2
	\; ,
	\\
	s^{-1} L^{-3}
	\; ,
	~~~~~ s \gtrsim L^2
	\; ,
\end{array}
\right.
\label{stoch_N}
\eea
with a crossover length $\ell_{\rm cross} \sim L^2$. We also note in Fig.~\ref{fig61}(b) that for very large $s$ there is a small bump, possibly corresponding to contributions from larger loops winding multiple times across the system.

\begin{figure}[t!]
	\centering
	\rotatebox{0}{\resizebox{.48\textwidth}{!}{\includegraphics{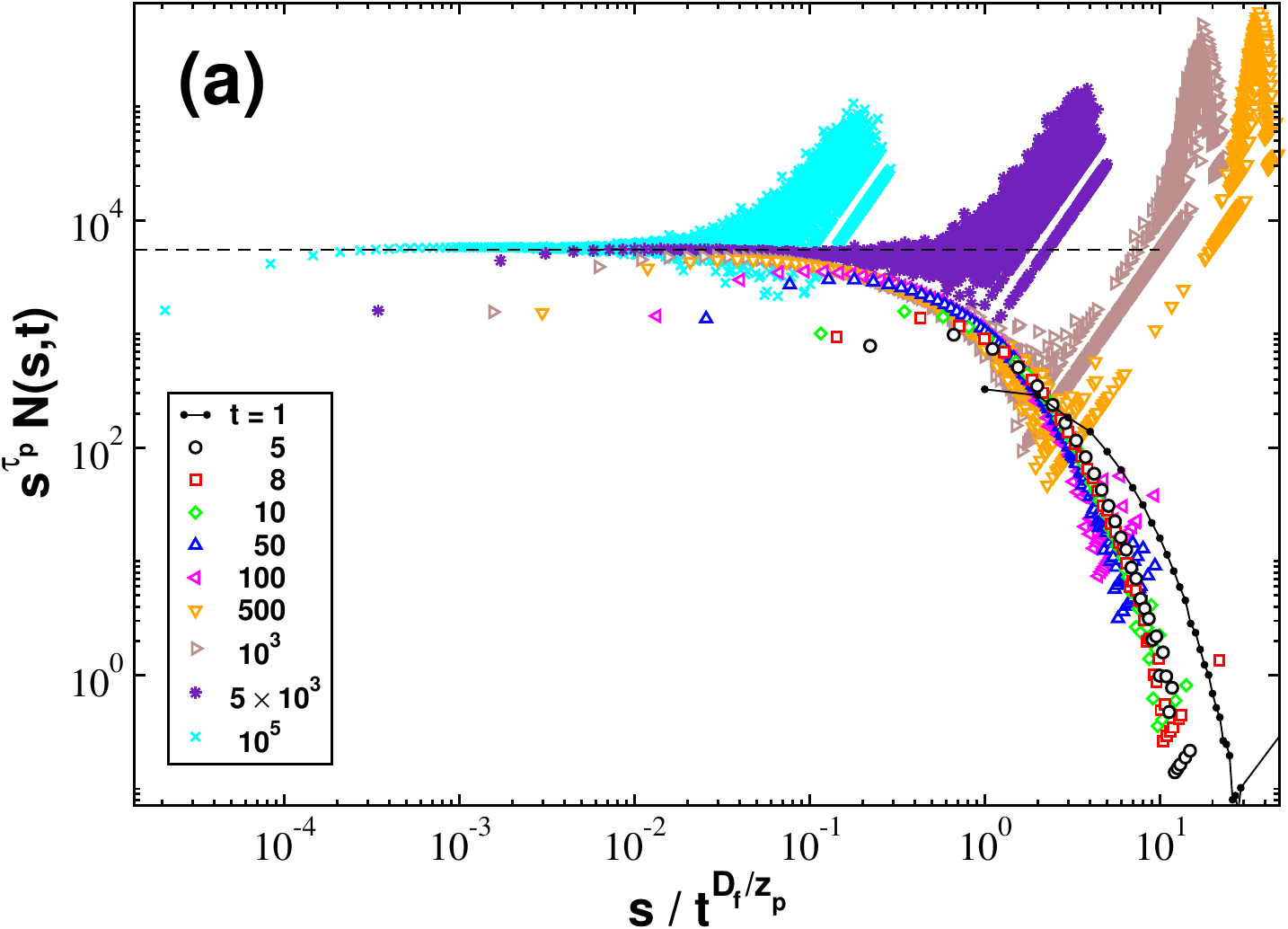}}}
	\rotatebox{0}{\resizebox{.48\textwidth}{!}{\includegraphics{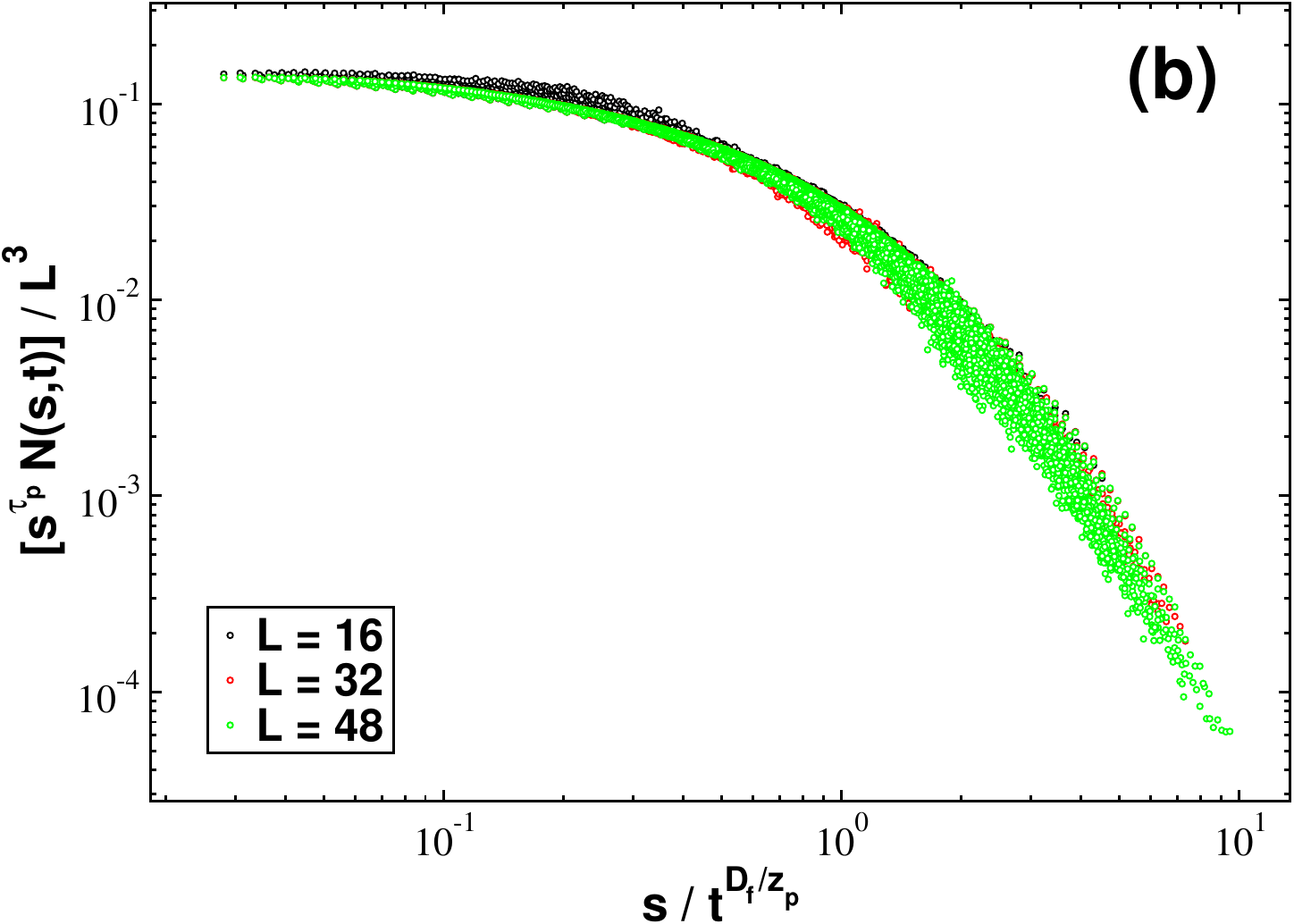}}}
	\caption{(a) Scaled variable $s^{\tau_{\rm p}} N(s,t)$ vs. $s/t^{D_{\rm f}/z_{\rm p}}$ (on a log-log scale) for FK clusters in a system of linear size $L=32$ quenched from $T= 2\, T_{\rm c}$ to $T_{\rm c}$. The horizontal line represents the law $N(s) \sim s^{-\tau_{\rm p}}$. Different datasets represent different timescales (see keys). (b) Same scaling plot, normalized by volume $L^3$, shown for different system sizes (see keys) and times in range $t \in [5,500]$, where data at different times are indicated by the same color. The plot is restricted to the scaling regime. In (a)-(b), $D_{\rm f} = 2.481$, $\tau_{\rm p} = 2.209$, and $z_{\rm p} = 2.65$.}
	\label{fig75}
\end{figure}

\subsection{Kinetics of FK clusters}

In this section, we discuss our numerical results concerning the growth kinetics of FK clusters following a critical quench from a high-temperature phase. For this purpose, we plot in Fig.~\ref{fig73} the number profile $N(s,t)$ of FK clusters versus their size $s$ for various times $t$ after the quench. In panel (a), the data are shown for different $t \in [1,500]$, while in (b) they are shown for later times up to $t = 10^5$. With increasing time (even as early as $t \simeq 5$), the distribution $N(s,t)$ changes significantly: the values of $N$ at small $s$ increase, while the bump at large $s$ shifts toward smaller $s$ and decreases in height. This indicates that large, system-spanning structures gradually shrink in both size and number, while smaller clusters become more abundant. Moreover, beyond a transient time $t_0 \sim 5$, a power-law decay in $N(s,t)$ of the form $N(s,t) \propto s^{-\tau_{\rm p}}$ is observed up to a cutoff size that increases with time. We find that the exponent $\tau_{\rm p}$ is consistent with its critical value, $\tau_{\rm p} \simeq 2.21$. At late times ($t > 10^3$ in panel (b)), the large-$s$ bump disappears and a single power-law regime persists, indicating that the system has reached critical equilibrium.

We note that, due to the probability-dependent inclusion of simple plaquettes in FK clusters, small-size clusters exhibit fluctuations that lead to deviations from the critical law~\eqref{power_law}. At very large $s$, finite-size effects also introduce corrections; see the data in panel~(b).

To check if a dynamical scale emerges during the quench dynamics, we propose the following scaling:
\be
N(s,t) \sim \frac{1}{s^{\tau_{\rm p}}} f_N \left( \frac{s}{[\xi_{\rm p}(t)]^{D_{\rm f}}} \right)
\; ,
\label{scale_N}
\ee
where $\tau_{\rm p}$ is the (critical) power-law exponent obtained above, and $D_{\rm f}$ is the fractal dimension of the critical structures, $D_{\rm f} = D/(\tau_{\rm p}-1)$. The time-dependent 
parameter $\xi_{\rm p}(t)$ is a \textit{putative} dynamical lengthscale, $\xi_{\rm p}(t) \propto t^{1/z_{\rm p}}$.

The scaling relation~\eqref{scale_N} can be tested by setting the parameters $\tau_{\rm p} = 2.209$ and $D_{\rm f} = 2.481$, while treating $z_{\rm p}$ as a free parameter to obtain the best data collapse. Because the number of objects of size $s$ has large fluctuations and its error bars are underestimated, the collapse goodness function $S$ (see Appendix~\ref{collapse}) does not reach a minimum value around $1$ during our analysis. However, the function $S$ exhibits a well-defined convex minimum at exponent $z_{\rm p} \simeq 2.65(8)$ across different scanning windows and system sizes.

In Fig.~\ref{fig75}(a), the data collapse is shown for a system of size $L=32$ with $z_{\rm p} = 2.65$. Let us discuss the scaling behavior observed in this figure. First, the bumps present in various datasets (also observed in Fig.~\ref{fig73}) arise from gigantic system-spanning clusters, which do not obey the scaling~\eqref{scale_N}, as expected. Second, as discussed earlier, small FK clusters exhibit fluctuations due to their construction, and therefore deviations from scaling at small $s$ are not unexpected (see also Ref.~\cite{33q3-g68k}). The remaining intermediate range of sizes $s$ follows the scaling form very well (as seen in the log-log plot) after a transient time $t \gtrsim 5$.

For concreteness, the scaling plots for $L=16, 32, 48$ in the intermediate range of sizes are shown in Fig.~\ref{fig75}(b). For each system size, various datasets at $t \in [5,500]$ are chosen. After normalization by the volume $L^3$, all datasets belonging to different $L$ and $t$ fall onto a common master curve, confirming the robustness of scaling function $f_N$ in Eq.~\eqref{scale_N}.

Thus, FK clusters support dynamical scaling in terms of a growing lengthscale $\xi_{\rm p}(t) \propto t^{1/z_{\rm p}}$ during the critical quench dynamics.

\subsection{Kinetics of geometrical line loops}

\begin{figure}[t!]
	\centering
	\rotatebox{0}{\resizebox{.99\textwidth}{!}{\includegraphics{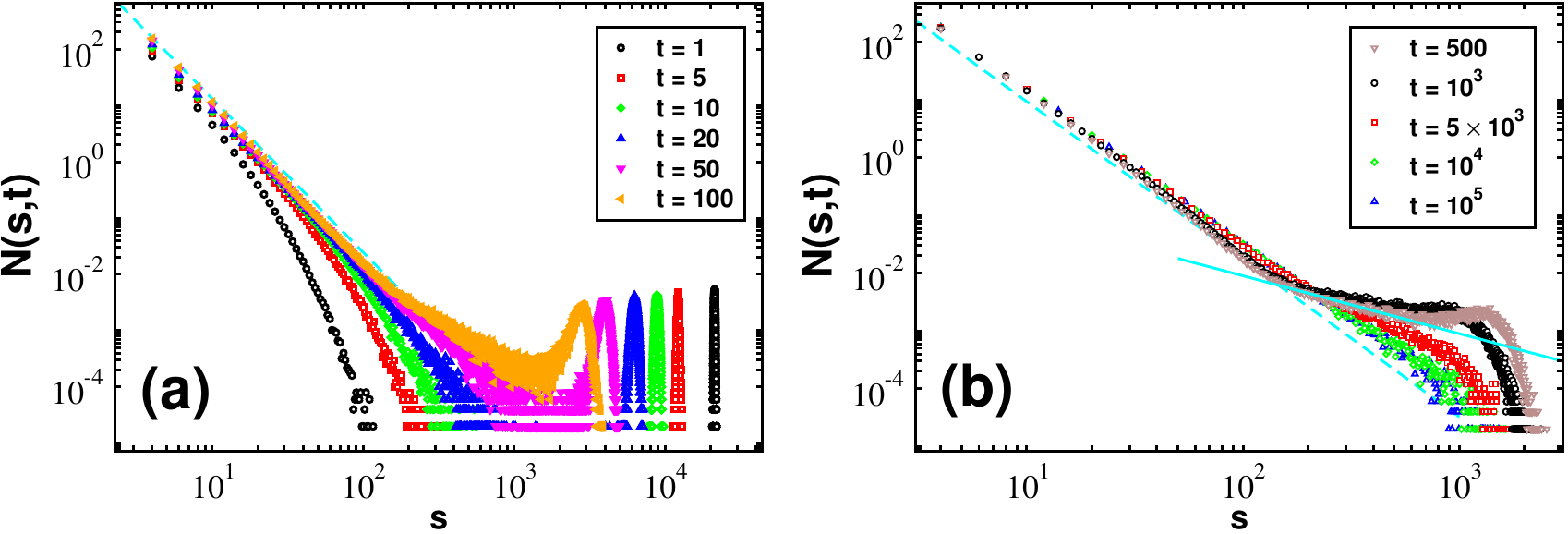}}}
	\rotatebox{0}{\resizebox{.99\textwidth}{!}{\includegraphics{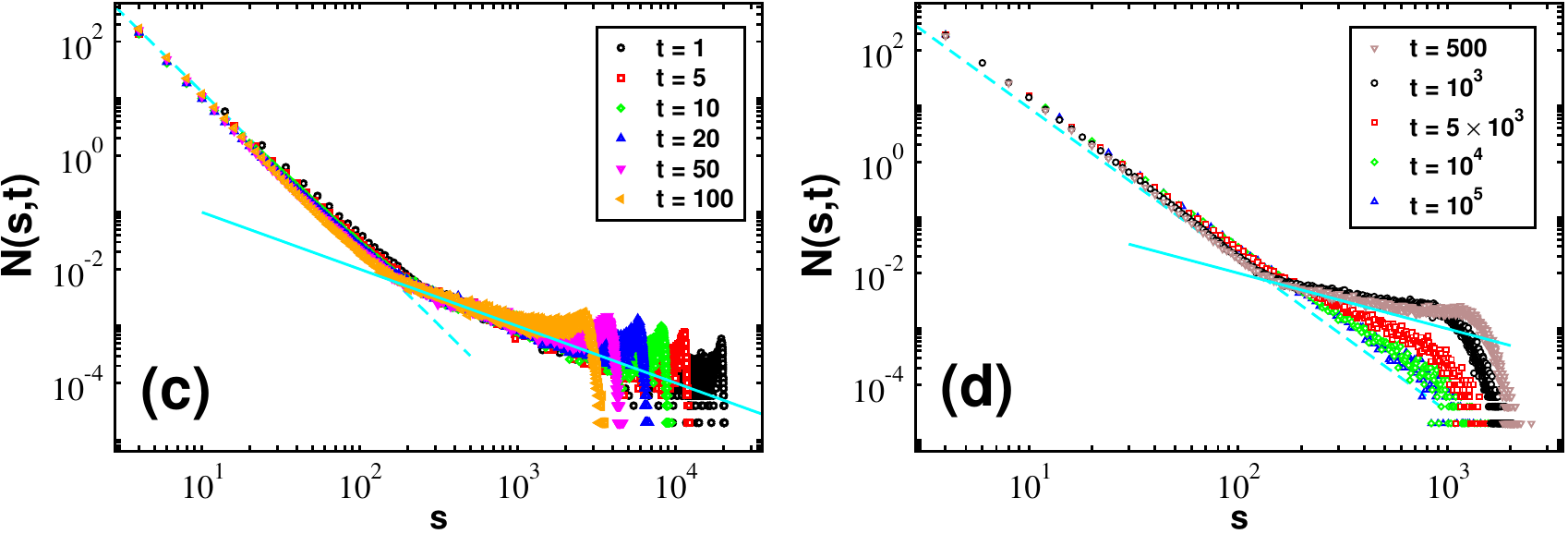}}}
	\caption{Number $N(s,t)$ vs. size $s$ on a log-log scale of loops constructed with (a)--(b) maximal rule, and (c)--(d) stochastic rule, for a system of linear size $L=32$ quenched from $T= 2\, T_{\rm c}$ to $T_{\rm c}$. Different datasets correspond to a different timescale (see keys) subsequent to the quench at time $t=0$. 
	The dashed line in all panels represents the power-law $N(s) \sim s^{-\tau}$ with $\tau = \tau_{\rm p}\,(= 2.729)$, while the solid line represents the linear fall with $\tau = 1$.
	}
	\label{fig71}
\end{figure}

In this section, we discuss the results for line loops constructed using the maximal and stochastic reconnection methods. In Fig.~\ref{fig71}, the number profiles $N(s,t)$ are shown for a quench from a high-temperature state to $T_{\rm c}$. In panels (a)--(b), the data for loops constructed using the maximal rule are shown, while in panels (c)--(d) the data for the stochastic rule are presented.

We first discuss the data in panels (c)--(d) for the stochastic method. As noted earlier, at temperatures above $T_{\rm c}$ these loops exhibit a two-fold power-law decay in addition to a bump at large $s$. This behavior is also observed during the initial time regime in panel (c). The peak shifts toward smaller $s$ with time, and the small-$s$ power law with exponent $\tau \simeq 2.5$ crosses over to the critical one ($\tau_{\rm p} \simeq 2.73$), which we discuss in more detail below. With increasing time, the linear decay of $N(s,t)$ at scales $s \gtrsim L^2$ also diminishes, and eventually, at late times ($t \gtrsim 5000$), only a single power-law regime with $\tau = \tau_{\rm p}$ survives.

\begin{figure}[t!]
	\centering
	\rotatebox{0}{\resizebox{.49\textwidth}{!}{\includegraphics{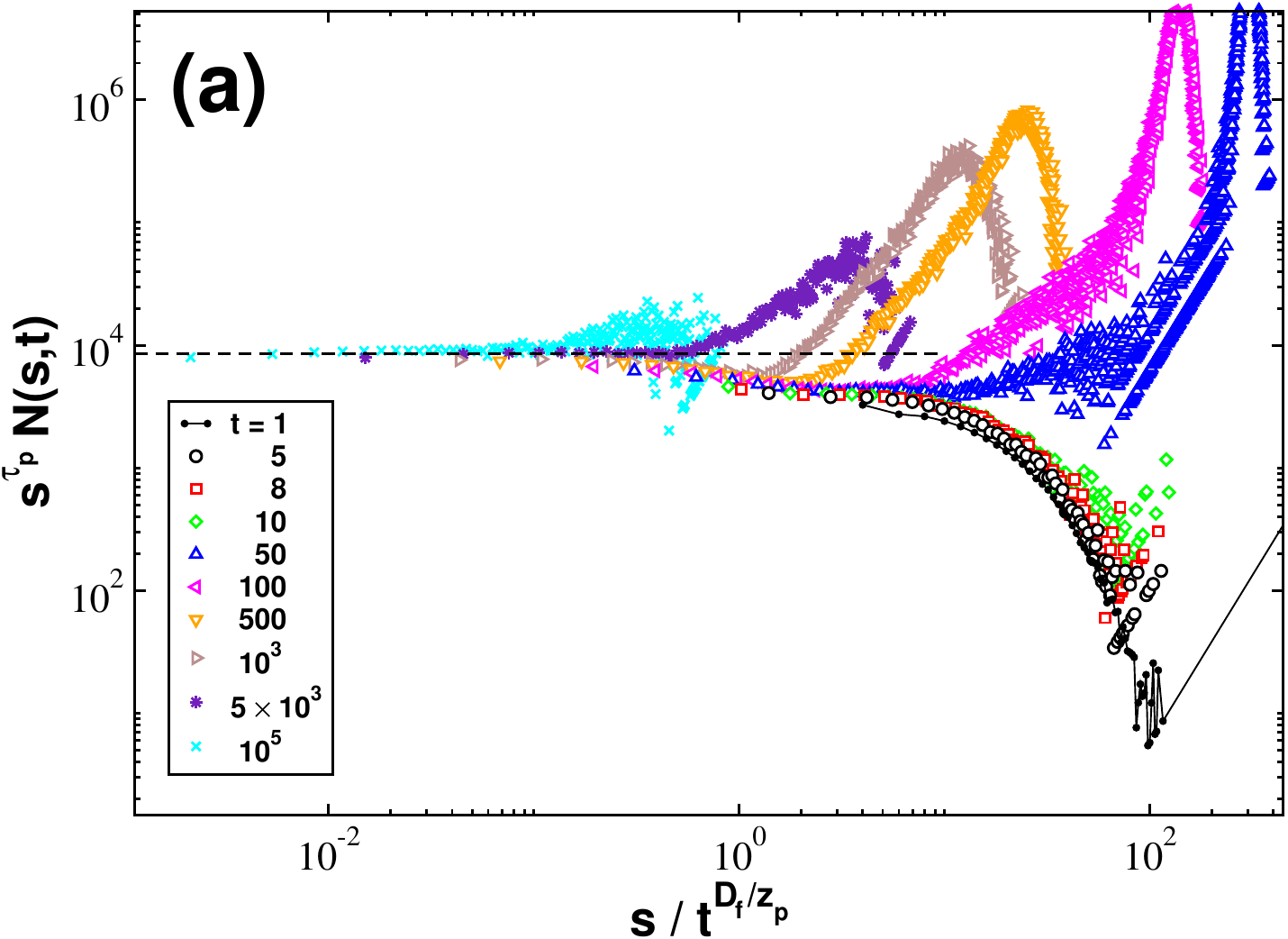}}}
	\rotatebox{0}{\resizebox{.49\textwidth}{!}{\includegraphics{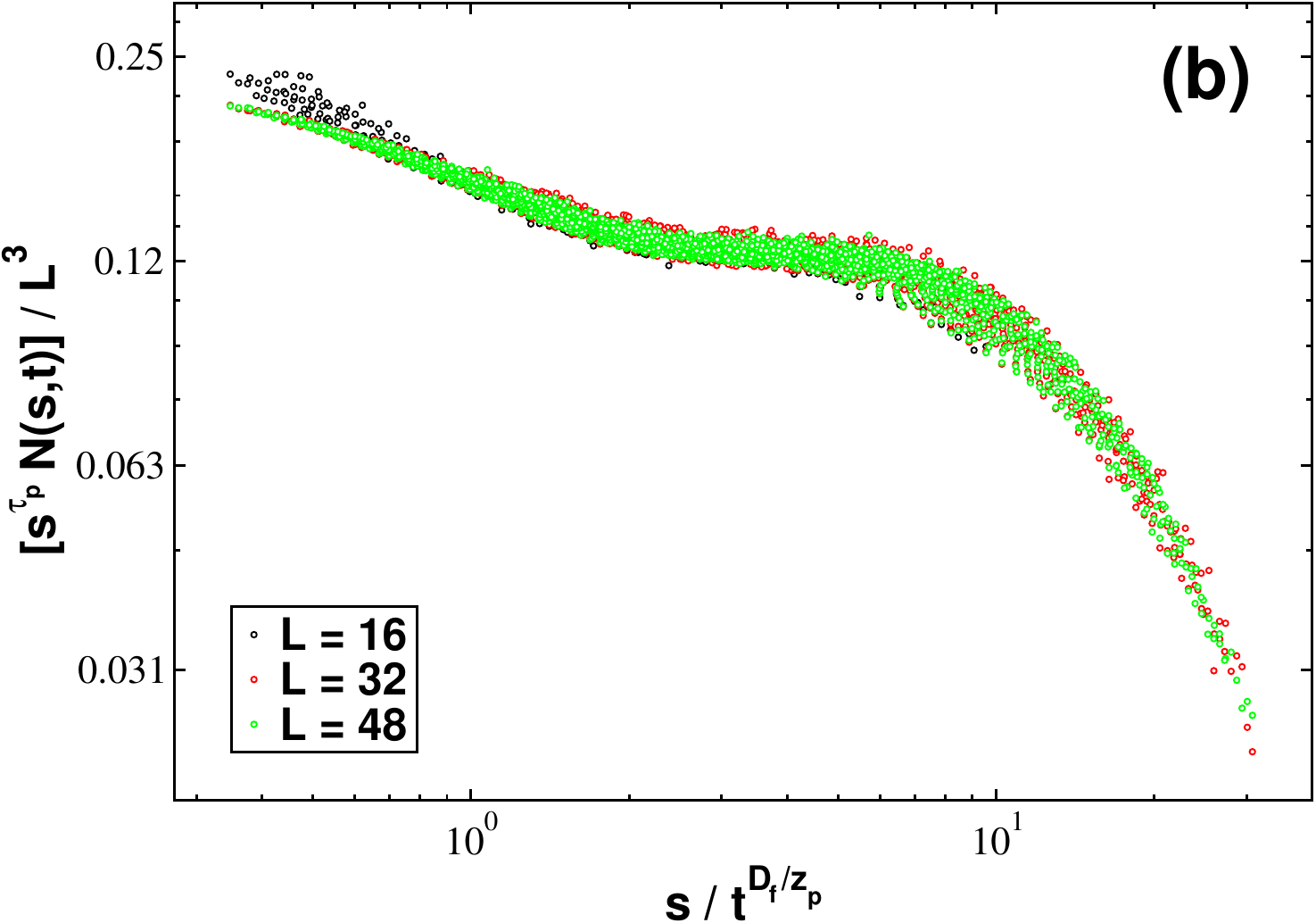}}}
	\rotatebox{0}{\resizebox{.49\textwidth}{!}{\includegraphics{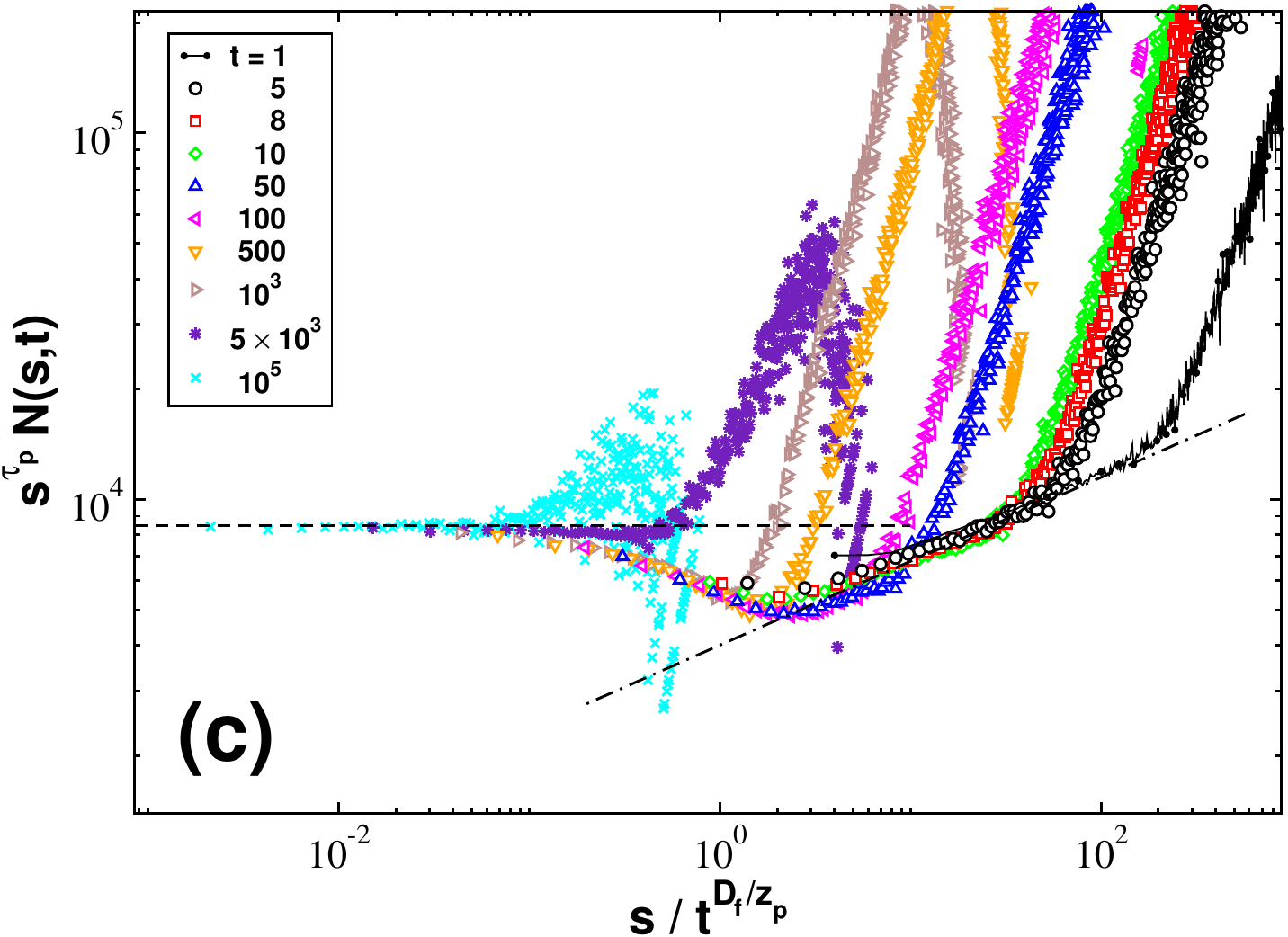}}}
	\rotatebox{0}{\resizebox{.49\textwidth}{!}{\includegraphics{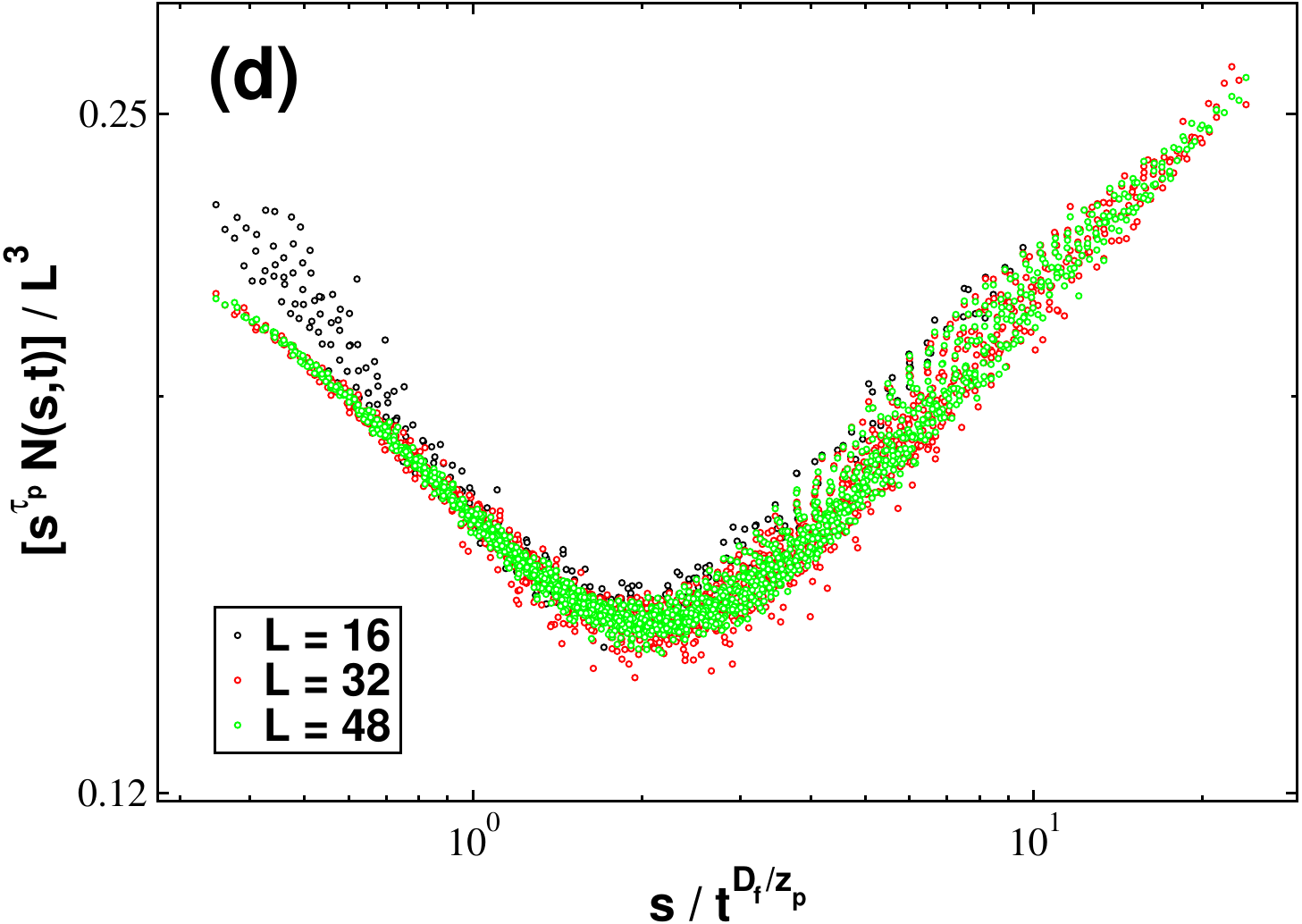}}}
	\caption{Plot of scaled variables $s^{\tau_{\rm p}} N(s,t)$ vs. $s/t^{D_{\rm f}/z_{\rm p}}$ (on a log-log scale) for loops constructed with (a) maximal rule, (c) stochastic rule, in a system of linear size $L=32$ quenched from $T= 2\, T_{\rm c}$ to $T_{\rm c}$. Different datasets represent different timescales (see keys). The horizontal dashed lines in both panels represent the power-law $N(s) \sim s^{-\tau_{\rm p}}$ with $\tau_{\rm p} = 2.729$, while the dashed-dot line in (c) represent the same law with $\tau = 2.5$. Panels (b) and (d), respectively, show scaling plots (normalized by volume $L^3$) for different system sizes (see keys) and times in range $t \in [5,500]$ (data at different times are indicated by the same color). The plots are restricted to the scaling regime. In all panels, $D_{\rm f} = 1.735$, $\tau_{\rm p} = 2.729$, and $z_{\rm p} = 2.6$.}
	\label{fig82}
\end{figure}

We now turn to the data for maximal reconnection in panels (a)--(b). Since these loops exhibit distinct behavior in the initial high-temperature state, their subsequent evolution is also different. During the initial stage of the evolution, the large system-spanning loops break up into many smaller loops, leading to an increase in the number of small- to intermediate-sized loops and a shift of the large-$s$ bump toward smaller $s$. This reflects the emergence of critical correlations, where geometrical structures of all sizes up to a growing lengthscale are present. Similar to the stochastic case, $N(s,t)$ for maximal reconnection exhibits a critical decay with $\tau_{\rm p} \simeq 2.73$ up to a cutoff size that increases with time.

Interestingly, just before reaching critical equilibrium ($t \gtrsim 5000$), when the large-$s$ bump has nearly disappeared, a new regime characterized by a linear decay, $N(s) \propto s^{-1}$, emerges at large $s$. Notably, for this reconnection rule, such a regime is absent in the initial high-temperature state due to the absence of non-contractible spanning loops~\cite{PhysRevD.49.4089} (although it is present for loops constructed using the stochastic rule). We will analyze this regime quantitatively below.

We now examine whether the dynamical scaling form~\eqref{scale_N} also holds during the evolution of line loops. For this purpose, we test the scaling function~\eqref{scale_N} by fixing the parameters $\tau_{\rm p} = 2.729$ and $D_{\rm f} = 1.735$ and treating exponent $z_{\rm p}$ as a free parameter in order to locate a well-defined minimum of the goodness function $S$~\eqref{goodness}. At fixed system size, we scan over different windows in the scaled size $s/t^{D_{\rm f}/z_{\rm p}}$ and find that, for both types of loops, the minimum value $S_{\rm min}$ is obtained for $z_{\rm p}$ in the broader range $2.2$--$2.65$.

In Fig.~\ref{fig82}, the scaling plots are shown for both types of loops with $z_{\rm p} = 2.6$. Let us first discuss the data in panel (c) for a system of size $L=32$ for stochastic reconnection. The curves exhibit two dynamical regimes: $N(s,t) \sim s^{-2.73}$ at small values (horizontal line) and $N(s,t) \sim s^{-2.5}$ (sloped dash-dotted line) at large values of the scaling variable $s/t^{D_{\rm f}/z_{\rm p}}$. With increasing time, a clear crossover to the critical law is observed. As far as the data collapse is concerned, apart from deviations at large $s$ where $N(s,t)$ shows a linear decay and a residual bump, a good scaling collapse is obtained over the remaining range of the data. A similarly good collapse is obtained for loops with maximal reconnection in panel (a). The portions of the datasets at large $s$ that deviate from scaling correspond to the unimportant bump structures and the newly emerged linear regime.

Furthermore, the scaling plots (after normalization) for various $L$ in the range of sizes where scaling~\eqref{scale_N} is valid are shown in panels (b) and (d) for loops constructed with maximal and stochastic rules, respectively. Again, in both panels, all datasets belonging to different $L$ and $t$ fall onto a single $L$-independent master curve. We therefore conclude that the line loops also support a dynamical scaling framework in terms of a growing lengthscale $\xi_{\rm p}(t) \propto t^{1/z_{\rm p}}$, although the determination of $z_{\rm p}$ remains somewhat noisy.

\begin{figure}[t!]
	\centering
	\rotatebox{0}{\resizebox{.7\textwidth}{!}{\includegraphics{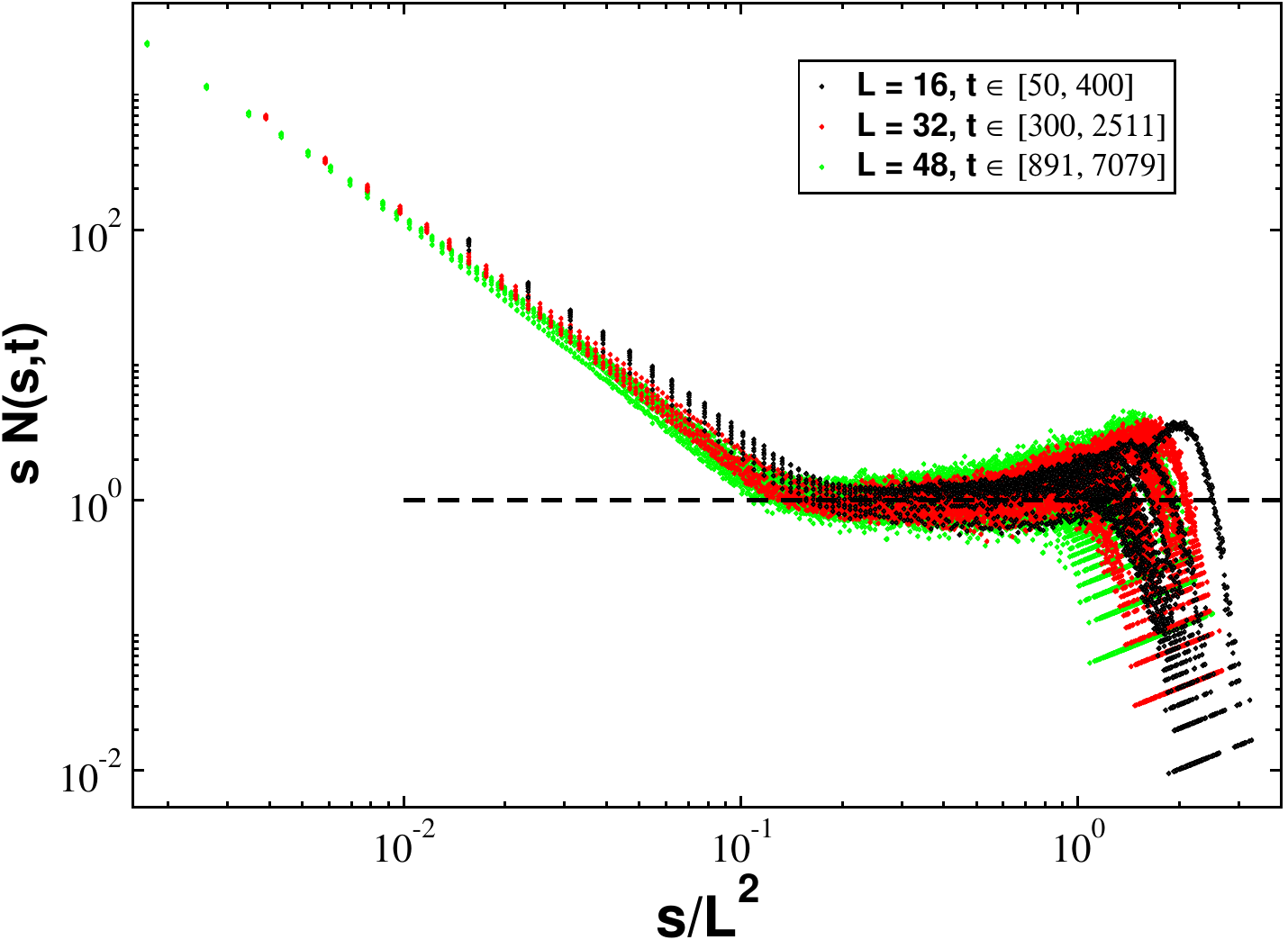}}}
	\caption{Scaling plot of $s N(s,t)$ vs. $s/L^2$ on a log-log scale for loops constructed with maximal rule in systems of linear size $L$ (see the keys) quenched from $T= 2\, T_{\rm c}$ to $T_{\rm c}$. The horizontal dashed line represents the linear fall, $N(s,t) \sim s^{-1}$.}
	\label{fig76}
\end{figure}

We now probe the linear decay observed in $N(s,t)$ in Fig.~\ref{fig71}(b). From Eq.~\eqref{stoch_N}, we know that the regime $N(s) \sim s^{-1}$ typically appears at scales $s \gtrsim L^2$, where the configuration becomes densely packed with large non-contractible loops. Motivated by this, we plot in Fig.~\ref{fig76} the scaled quantity $s\,N(s,t)$ versus $s/L^2$ for loops constructed using the maximal rule, for different system sizes ($L=16,32,48$) and at various times $t$ after a quench from high temperature to $T_{\rm c}$. The time window $t \in [t_{\rm min}, t_{\rm max}]$ for each system size is chosen such that $N(s,t)$ exhibits a linear decay at large $s$. The plateau observed in such a plot clearly confirms the dynamical regime $N(s,t) \sim s^{-1}$. Interestingly, the extremal values of these time windows follow the scaling relation $t \propto L^{z_{\rm p}}$.

Such a dynamical regime has also been observed for vortex tangles formed using the same maximal reconnection rule in $\mathcal{U}(1)$ complex field theories~\cite{PhysRevE.94.062146,Kobayashi_2016} during quenches below the critical temperature. A possible explanation for its emergence is that, during the dynamics, large loops proliferate at the expense of multiple spanning ones (which are responsible for the residual bump) and become non-contractible, irrespective of the initial state in which only a few gigantic but contractible loops are present.

\section{Conclusions}
\label{S5}

We investigated the critical dynamics of Wegner's three-dimensional  $\mathbb{Z}_2$ lattice gauge model after instantaneous quenches from the high temperature phase and the zero temperature ground state
to the critical point $T_{\rm c}$, where a transition of topological origin with no local order parameter takes place. The equilibrium critical behavior of the model is more naturally described in terms of the percolation of geometrically defined objects~\cite{33q3-g68k}, namely line loops and Fortuin--Kasteleyn (FK) clusters. We exploited this geometrical framework to understand the dynamical critical behavior of the model.

Through an extensive analysis of time-dependent finite-size scaling functions for the relaxation of both the energy density and the percolation order parameter, we determined the dynamical critical exponents $z_{\rm c}$ and $z_{\rm p}$, respectively. We found that the nonequilibrium relaxation of both thermal and geometrical observables is governed by a single exponent, with the estimate $z_{\rm p} \simeq 2.6$ being consistent with the value of $z_{\rm c}$. Thus, using geometrical observables, we confirmed that the dynamical exponent in the $\mathbb{Z}_2$ gauge model is distinct from that of the Ising universality class ($z_{\rm c} \simeq 2.02$). This is not surprising, as the Kramers--Wannier (KW) duality maps the high-temperature expansion of the partition function of the 
Ising model to the low-temperature expansion of its dual model, rather than relating the corresponding local update algorithms.

We also explored the morphology of different geometrical objects during the critical quench dynamics from a high-temperature percolation phase. Since the $\mathbb{Z}_2$ gauge model lacks a local order parameter and is instead characterized by global ones, it is not \textit{a priori} clear whether a dynamical lengthscale analogous to that in systems with local order parameters emerges in this system. We addressed this question from a geometrical perspective by examining the growth kinetics of these objects.

We found that the number statistics $N(s,t)$ of both FK clusters and line loops (constructed using maximal and stochastic reconnection rules) support the dynamical scaling framework in terms of a time-dependent lengthscale, $\xi_{\rm p}(t) \sim t^{1/z_{\rm p}}$. Moreover, the evolution of these geometrical objects exhibits several interesting dynamical regimes:
\begin{enumerate}[label=\arabic*., nosep]
\item
At early times, large multiple-spanning objects shrink so that smaller ones can accommodate.
\item
Small-sized objects progressively reorganize into structures with critical characteristics, such that the number statistics $N(s,t)$ at small $s$ crosses over to a critical form $N(s,t) \sim s^{-\tau}$, with $\tau \simeq 2.21$ for FK clusters and $\tau \simeq 2.73$ for line loops.
\item
Large line loops, irrespective of the reconnection rule, resemble the statistics of fully packed loop models~\cite{PhysRevD.49.4089,PhysRevLett.107.177202,PhysRevLett.111.100601}.
\end{enumerate}

Finally, we demonstrated the usefulness of percolation-based observables in understanding dynamical critical phenomena and we provided new insights into systems with global order parameters, including gauge-theoretic models. As a future direction, an extension of the present framework to disordered gauge theories is currently underway~\cite{Agrawal_inprep}.

\vspace{0.5cm}

\noindent
{\bf Acknowledgements.}
R.A. acknowledges support from the Bando Ricerca Scientifica 2025--Avvio alla Ricerca (D.R. No. 2155/2025) of Sapienza Università di Roma, project  B83C25004300005--OptiRFVM. The simulations were performed on the SACADO MeSU platform at Sorbonne Universit\'e Paris. R.A. thanks A. Pelissetto for useful conversation. We also thank L. Faoro and L. Ioffe for early discussions on this model.

\vspace{0.5cm}

\appendix
\renewcommand{\thesection}{\Alph{section}}

\section{Goodness measure for data collapse}
\label{collapse}

Obtaining precise values of free variables (with correct error bounds) from the optimization of a data collapse is non-trivial. For a sensitive analysis, we exploit a method developed by Houdayer and Hartmann~\cite{PhysRevB.70.014418} (also see Ref.~\cite{doi:10.1143/JPSJ.62.435}). For scaled data points $\{(x_{ij},{y}_{ij},\Delta{y}_{ij})\}$ participating in the collapse, this method defines a reduced $\chi^2$-like goodness measure,
\be
S = \frac{1}{N_{\text{dof}}}\sum_{ij} \frac{\bigl({y}_{ij} - Y_{ij}\bigr)^2}{\Delta{y}_{ij}^{\,2} + \sigma_{Y_{ij}}^2}
\; ,
\label{goodness}
\ee
where $Y_{ij}$ and $\sigma^2_{Y_{ij}}$ are the value and the variance of a \textit{single} master curve at point $x_{ij}$, and $N_{\text{dof}}$ is the number of total data points. Master-curve values $Y_{ij}$ and $\sigma^2_{Y_{ij}}$ are estimated as follows: For a given point $(x_{ij},{y}_{ij},\Delta{y}_{ij})$ belonging to a system of size $L_i$, we examine datasets from all other system sizes $L_{i'} \neq L_i$. For each such $L_{i'}$, we collect the two nearest neighbors in $x$ that bracket $x_{ij}$: $x_{{i'},\ell} < x_{ij}$ and $x_{{i'},r} > x_{ij}$. The collected set is then used to perform a weighted linear fit with weights $w = 1/(\Delta{y})^2$. This fit yields the predicted master‑curve value $Y_{ij}(x_{ij})$ and its variance $\sigma_{Y_{ij}}^2(x_{ij})$.

In general, $S$ is a convex function of free variables (dynamical exponent in our case), and the goodness of a data collapse is quantified by an optimal value $S = S_{\rm min} (\simeq 1)$. We estimate the error bounds on the extracted exponent by accepting $S_{\rm min} + 1.5$ as goodness criterion. In practice, achieving values in this range requires a careful iterative procedure --- we scan through various ranges of scaled abscissa, adjust the free variable, and systematically exclude small system sizes that exhibit significant sub-leading corrections. Consequently, the number of data points entering the goodness measure, $N_{\rm dof}$, varies from one analysis to another depending on the selected scaling windows and the system sizes retained. Typically, between $40$ and $120$ data points contribute to each analysis.

\addcontentsline{toc}{section}{References}
\bibliographystyle{iopart-num}
\bibliography{ref.bib}

\end{document}